\documentclass[%
 aip,jcp,
 amsmath,amssymb,
 reprint,
]{revtex4-2}

\usepackage[a-1b]{pdfx}

\usepackage{dsfont}

\usepackage{physics}
\usepackage{graphicx}% Include figure files
\usepackage{dcolumn}% Align table columns on decimal point
\usepackage{bm}% bold math
\usepackage[utf8]{inputenc}
\usepackage[T1]{fontenc}
\usepackage{mathptmx}
\usepackage{listings}
\usepackage{rotating} % Rotating table

\usepackage{color}
\usepackage{dcolumn} % decimal align in tables
\usepackage{bm} % bold math
\usepackage{graphicx}
\usepackage{multirow} % for table cells to span rows
\usepackage{pifont} % for checkmarks
\usepackage{epsfig}
\usepackage{amsmath} % matrix
\usepackage{subfigure}
\usepackage{float}
\usepackage{booktabs}
\usepackage{tabularx}
\usepackage{natbib}
\usepackage{gensymb}
\usepackage{rotating}
\usepackage{threeparttable}
\usepackage{comment}
\usepackage[normalem]{ulem}
\usepackage{etoolbox}
\makeatletter
\newcommand{\maybecompressarraystretch}{%
  \@ifclasswith{revtex4-2}{preprint}{%
    \renewcommand\arraystretch{0.85}%
  }{}%
}
\makeatother
\begin{document}
  
\author{Run R. Li}
\affiliation{
             Department of Chemistry and Biochemistry,
             Florida State University,
             Tallahassee, FL 32306-4390, USA}          

\author{Stephen H. Yuwono }
\affiliation{
             Department of Chemistry and Biochemistry,
             Florida State University,
             Tallahassee, FL 32306-4390, USA}          
\author{Marcus D. Liebenthal }
\affiliation{
             Department of Chemistry and Biochemistry,
             Florida State University,
             Tallahassee, FL 32306-4390, USA}     

\author{Tianyuan Zhang}
\affiliation{Department of Chemistry, University of Washington, Seattle, WA 98195, USA}

\author{Xiaosong Li}
\affiliation{Department of Chemistry, University of Washington, Seattle, WA 98195, USA}
             
\author{A. Eugene DePrince III}
\email{adeprince@fsu.edu}
\affiliation{
             Department of Chemistry and Biochemistry,
             Florida State University,
             Tallahassee, FL 32306-4390, USA}

\title{Relativistic Two-component Double Ionization Potential Equation-of-Motion Coupled Cluster with the Dirac--Coulomb--Breit Hamiltonian}

\begin{abstract}
We present an implementation of relativistic double-ionization-potential (DIP) equation-of-motion coupled-cluster (EOMCC) with up to 4-hole--2-particle (4$h$2$p$) excitations that makes use of the molecular mean-field exact two-component (mmfX2C) framework. We apply mmfX2C-DIP-EOMCC to several neutral atoms and diatomic molecules to obtain the ground and first few excited states of the corresponding di-cation species, and we observe excellent agreement (to within 0.001 eV) between double ionization potentials (IPs) obtained from mmfX2C- and four-component DIP-EOMCC calculations that include 3-hole--1-particle ($3h1p$) excitations, with either the Dirac--Coulomb or Dirac--Coulomb--Gaunt Hamiltonians. We also compare double IPs for mmfX2C-DIP-EOMCC calculations with the full Dirac--Coulomb--Breit (DCB) Hamiltonian to those from experiment. The mmfX2C-DIP-EOMCC with $3h1p$ excitations leads to errors in absolute double IPs that
{ generally overestimate experimental data for noble gases by 0.1--0.4 eV,
whereas the inclusion of $4h2p$ excitations results in double IPs that are too low by 0.1--0.2 eV, at the large basis set limit.}

\end{abstract}

\maketitle

\section{Introduction}

\label{SEC:INTRODUCTION}

High-accuracy simulations of ground-state electronic structure often rely on the coupled-cluster (CC) family of methods.\cite{Coester58_421,Kuemmel60_477,Cizek66_4256,Cizek69_35,Shavitt72_50,Li99_1,Musial07_291} The popularity of CC approaches stems from their systematic improvability and rapid convergence to the full configuration interaction (CI) limit [from CC with single and double excitations (CCSD)\cite{Bartlett82_1910,Zerner82_4088} to CCSD plus triple excitations (CCSDT)\cite{Bartlett87_7041,Schaefer88_382} and beyond], as well as the size-extensivity and separability of the CC energy, regardless of the level at which the cluster operator is truncated.   For the description of electronically excited states, CC theory can be extended using various approaches that inherit the desirable properties of CC, including equation-of-motion (EOM)\cite{Emrich81_379,Bartlett89_57,Bartlett93_7029} or linear response (LR) CC,\cite{Monkhorst77_421,Monkhorst83_1217,Mukherjee79_325,Bartlett84_255,Paldus86_1486,Jorgensen90_3333,Helgaker90_3345} as well as the related symmetry-adapted cluster (SAC) CI framework.\cite{Nakatsuji79_334} The strengths of these approaches aside, certain situations remain challenging, such as the description of systems that exhibit multi-reference (MR) character, examples of which include open-shell radicals and diradicals, potential energy curves along bond-breaking coordinates, transition metal complexes with multiple unpaired electrons, and excited states that are dominated by two-electron transitions.

Multi-reference CC approaches seem to be the natural choice to treat systems with MR character, but they come with various complications such as intruder states, size extensivity issues, and the ambiguity in the specification of the MRCC wave function \emph{ansatz} itself (see, \emph{e.g.}, Refs.~\citenum{Li99_1,Kowalski02_676,Musial07_291,Bartlett12_182,Evangelista18_030901} for reviews of MRCC approaches). On the other hand, a number of strategies from the single-reference (SR) CC realm can be used to build wave functions for systems with MR character. These approaches could be as direct as the application of CC/EOMCC on top of a broken-symmetry Hartree--Fock (HF) reference function or the use of spin-flip CC/EOMCC,\cite{Head-Gordon09_044103,Krylov01_375,Krylov06_83,Krylov08_433} which exploits the mostly SR nature of a high-spin HF configuration. While straightforward, such strategies may be problematic as the resulting wave functions can exhibit significant spin contamination. An alternative approach is given by the use of particle-non-conserving operators that add or remove electrons from a closed-shell $N$-electron state that is well-described by SR CC; examples in this domain include the ionization potential (IP)\cite{Stanton94_65,Snijders92_55,Snijders93_15,Gauss94_8938,Bartlett03_1128,Bartlett04_210,Gauss05_154107,Wloch05_134113,Wloch06_2854,Piecuch06_234107} or electron attachment (EA)\cite{Bartlett95_3629,Bartlett95_6735,Bartlett03_1901,Wloch05_134113,Wloch06_2854,Piecuch06_234107} EOMCC, which take inspiration from Fock-space CC methods. If the $N$-electron state is described using SR CC built on top of a restricted HF (RHF) configuration, the open-shell wave function remains spin-adapted, and the expected degeneracy structure will be conserved. In this work, we explore the double IP (DIP)\cite{Nooijen02_65,Nooijen02_656,Bartlett11_114108,Krylov11_084109,Krylov12_244109,Piecuch13_194102,Piecuch14_868,Piecuch25_061101} EOMCC method, which allows, for example, the description of $p^4$ or $d^8$ electron configurations generated from the corresponding $p^6$ or $d^{10}$ closed-shell references, respectively, as well as direct determination of double ionization energies.

In addition to the accurate treatment of electronic correlation effects, quantitative predictions involving open-shell species may also require a sophisticated treatment of spin-free and spin-dependent relativistic effects, particularly when studying spin--orbit coupling dependent phenomena. For these purposes, the four-component (4c) Dirac--Coulomb--Breit (DCB) Hamiltonian offers the most complete description of relativistic effects in fermionic systems. However, the DCB Hamiltonian contains both the electronic and positronic degrees of freedom, the latter of which are not directly relevant to quantum chemistry applications and lead to unnecessary increases in the cost and complexity of simulations on large molecular systems. As such, approximations that effectively decouple the electronic and positronic degrees of freedom are desirable.  { Along these lines, two-component (2c) transformation approaches, which downfold the  relativistic physics in the 4c space into a smaller 2c space without much loss in accuracy, 
have gained popularity in the past couple of decades. Some such methods include the normalized elimination of the small component (NESC),\cite{Dyall97_9618,Dyall98_4201,Enevoldsen99_10000,Dyall01_9136,Cremer02_259} quasi-four-component (Q4C),\cite{Liu05_241102,Peng06_044102,Cheng07_104106} infinite-order two-component (IOTC),\cite{Saue07_064102} and exact two-component (X2C)\cite{Peng09_031104,Ilias09_124116,Liu10_1679,Liu12_154114,Reiher13_184105,Visscher14_041107,Li16_3711,Li16_104107,Repisky16_5823,Li17_2591,Gomes18_174113,Cheng18_034106,Cheng19_074102,Visscher21_5509,Cheng21_e1536,Li22_2947,Li22_2983,Li22_5011,Li24_3408,Li24_7694,Li24_041404} methods. }
{ This work focuses on the X2C family, which includes one-electron X2C (1eX2C),\cite{Peng09_031104,Li16_3711,Li16_104107,Repisky16_5823,Li17_2591,Li22_2947,Li22_2983,Li22_5011,Li24_7694,Li24_041404} atomic mean field X2C (amfX2C)\cite{Cheng18_034106,Cheng19_074102,Visscher21_5509,Cheng21_e1536} and molecular mean-field X2C (mmfX2C)\cite{Ilias09_124116,Visscher14_041107,Gomes18_174113,Li24_3408} variants that differ in the treatment of two-electron spin--orbit interactions. The 1eX2C approach either ignores these effects altogether or approximates them via an empirical scaling parameter.\cite{Boettger00_7809} The latter two approaches incorporate these effects at the mean-field level at different levels of sophistication. The amfX2C approach is a local approximation to mmfX2C that reduces the cost of the mean-field step.  Computationally, the difference between these approaches lies in solving 4c Hartree--Fock (4cHF) equation, which scales $\mathcal{O}(N_{\rm basis}^4)$, for each individual atom (amfX2C) or the whole molecule (mmfX2C). We advocate mmfX2C when applied in conjunction with the CC family of methods, because the computational cost of the mmfX2C transformation is small compared to that of the CC iterations, which scale $\mathcal{O}(N_{\rm basis}^6)$ or higher, depending on the CC approximation. In addition, recent computational developments\cite{Li21_3388, Li22_064112} have significantly reduced the cost of 4cHF method itself.
}

Several applications of DIP-EOMCC can be found in the literature that incorporate relativistic effects through the { 2c} framework or other means. For example, Wang and coworkers\cite{Guo15_144109, wang20_134105}  used effective core potential (ECP) to include the scalar relativistic effect at the mean-field level and applied a one-electron relativistic Hamiltonian in the post-HF treatment.  More recently, Piecuch and coworkers\cite{Piecuch25_061101} have implemented DIP-EOMCCSDT with up to 4-hole--2-particle ($4h2p$) excitations included in the EOMCC operator on top of the CCSDT ground-state reference, as well as an approximate form of this method, in combination with a spin-free relativistic Hamiltonian. Probably the most sophisticated treatment of relativistic effects in this context has been provided by Pathak and coworkers,\cite{Pal14_062501, Pal20_104302}  who developed 4c DIP-EOM-CCSD with the Dirac--Coulomb (DC) and Dirac--Coulomb--Gaunt (DCG) Hamiltonians at both the HF and post-HF levels; these methods were subsequently used to calculate the DIPs of alkaline metal atoms, rare gas atoms, and a few diatomic molecules. In the present work, we push the correlation and relativistic treatments in DIP-EOMCC to include both high-order correlation effects ({\em i.e.}, DIP-EOMCCSDT with $4h2p$ transitions) and the full DCB Hamiltonian within the molecular mean-field X2C (mmfX2C) approach.\cite{Ilias09_124116,Visscher14_041107,Gomes18_174113,Li24_3408} %

The remainder of this paper is organized as follows: Section \ref{SEC:THEORY} provides the relevant details of the X2C relativistic framework and CC/DIP-EOMCC correlation treatment.  The details of our computations are presented in Sec. \ref{SEC:COMPUTATIONAL_DETAILS}, after which the results of these computations are discussed in Section \ref{SEC:RESULTS}. Lastly, Sec.~\ref{SEC:CONCLUSIONS} provides some concluding remarks. 

\section{Theory}

\label{SEC:THEORY}

\subsection{The X2C framework}

The relativistic many-electron Hamiltonian takes the form
\begin{align}
\label{eqn:hamiltonian}
    &\hat{H} = \sum_{i} \left [ (\beta - \mathds{1}_4) mc^2 + c(\alpha_i \cdot \mathbf{\hat{p}}_i) + \sum_{A} \hat{V}_{iA}\right ] \nonumber \\
    & + \sum_{i<j} \left [ \frac{1}{r_{ij}} - \frac{\bm{\alpha}_i\cdot\bm{\alpha}_j}{r_{ij}} + \frac{1}{2} \left( \frac{\bm{\alpha}_i\cdot\bm{\alpha}_j}{r_{ij}} - \frac{\bm{\alpha}_i\cdot\mathbf{r}_{ij}\bm{\alpha}_j\cdot\mathbf{r}_{ij}}{r_{ij}^3} \right) \right ]
\end{align}
where $m$ is mass of the electron, $c$ is the speed of light, $\mathds{1}_4$ is the 4c identity matrix, $\mathbf{\hat{p}}_i$ is the momentum operator for electron $i$, $\bm{\alpha}_i$ and $\bm{\beta}$ are the Dirac matrices, $\hat{V}_{iA}$ is the potential energy operator for electron $i$ in the field of nucleus $A$, and $r_{ij}$ represents the distance between electrons $i$ and $j$.  The three two-electron terms in Eq.~\ref{eqn:hamiltonian} are referred to as the Coulomb, Gaunt, and gauge terms. When accounting for all three, we refer to $\hat{H}$ as the full Dirac--Coulomb--Breit (DCB) Hamiltonian. Approximations to the DCB Hamiltonian include the Dirac--Coulomb (DC) Hamiltonian, which accounts for only the Coulomb term ($r_{ij}^{-1} \mathds{1}_4$), and the Dirac--Coulomb--Gaunt (DCG) Hamiltonian, which accounts for the Coulomb term and the Gaunt term [$- \bm{\alpha}_i\cdot\bm{\alpha}_j (r_{ij})^{-1}$].

In the mmfX2C approach, one begins with a 4c-HF calculation carried out with the DC, DCG, or DCB relativistic Hamiltonian. The electronic molecular spinors obtained via this procedure define the unitary 4c to 2c transformation, which is applied to the Fock operator and the Coulomb part of the two-electron operator. The subsequent correlation treatment is then carried out with these operators in the transformed 2c basis. This procedure ensures that two-electron relativistic effects are captured self-consistently at the mean-field level, while it is assumed that correlation effects stemming from the Gaunt or gauge terms will be small and can be neglected in the post-HF part of the calculation. In this work, the notations DC-/DCG-/DCB-X2C refer to mmfX2C-based calculations employing the appropriate relativistic Hamiltonians.

In addition to the mmfX2C approach, we also consider the simpler one-electron X2C (1eX2C) in which a unitary 4c to 2c transformation is applied directly to the one-body part in Eq.~\ref{eqn:hamiltonian} (the core Hamiltonian). The HF and correlation treatments are carried then out in the resulting 2c basis, with a non-relativistic Coulomb operator. The missing two-body spin--orbit interaction effects can be approximated by scaling the spin--orbit part of the core Hamiltonian using a screened nuclear spin--orbit (SNSO) factor.\cite{Boettger00_7809} In this work, the 1eX2C procedure uses the row-dependent factors parametrized for the DCB Hamiltonian in Ref.~\citenum{Li23_5785}. For additional detailed comparisons of the mmfX2C and 1eX2C approaches, we refer the reader to Refs.~\citenum{Li24_3408}, \citenum{Ilias09_124116}, \citenum{DePrince25_084110}, and the references contained therein.

\subsection{DIP-EOMCC Theory}

In this section, we provide the pertinent details of the CC and DIP-EOMCC approaches, the latter of which is used to model doubly ionized states. Throughout the discussion, the labels $i_1, i_2, \ldots$ and $a_1, a_2, \ldots$ refer to molecular spinors that are occupied and unoccupied in the reference (1eX2C- or mmfX2C-HF) configuration. We also make use of the Einstein summation convention, where repeated lower and upper indices imply summation. 

In the CC approach, the ground-state electronic wave function for an $N$-electron system is expanded as 
\begin{align}
| \Psi_0^{(N)}\rangle = \exp(\hat{T})|\Phi_0 \rangle
\end{align}
where $|\Phi_0\rangle$ is the reference configuration. The cluster operator, $\hat{T}$, is expanded in terms of products of particle-conserving excitation operators, {\em i.e.}, 
\begin{equation}
\label{EQN:CC}
    \hat{T} = \sum_{n=1}^{M} \hat{T}_n,\;
    \hat{T}_n = \left(\frac{1}{n!}\right)^2 t_{a_1 \ldots a_n}^{i_1 \ldots i_n} \prod_{k=1}^{n} (\hat{a}^{a_k} \hat{a}_{i_k})
\end{equation}
where $t_{a_1 \ldots a_n}^{i_1 \ldots i_n}$ are the cluster amplitudes, and the symbols $\hat{a}^{a_k}$ and $\hat{a}_{i_k}$ represent creation and annihilation operators for molecular spinors $a_k$ and $a_i$, respectively. Here, the parameter $M$ determines the level in the CC hierarchy of methods, and the full CC ($\equiv$ full CI) limit is reached when $M = N$ (where $N$ is the number of electrons). In this work, we are concerned with the $M=2$ (CCSD) and $M=3$ (CCSDT) levels. The cluster amplitudes are determined in the usual projective way, {\em i.e.}, by solving 
\begin{equation}
    \label{EQN:CC_amplitudes}
        \mel*{\Phi_{i_1 \ldots i_n}^{a_1 \ldots a_n}}{\bar{H}}{\Phi_0} = 0 \; \forall \; \ket*{\Phi_{i_1 \ldots i_n}^{a_1 \ldots a_n}}, n = 1, 2, ..., M.
\end{equation} 
where we have introduced the similarity-transformed Hamiltonian, $\bar{H} = e^{-\hat{T}} \hat{H} e^{\hat{T}}$. In Eq.~\ref{EQN:CC_amplitudes}, the symbol $\ket*{\Phi_{i_1 \ldots i_n}^{a_1 \ldots a_n}}$ refers to a determinant of spinors that is $n$-tuply substituted relative to the reference determinant. Once the amplitudes have been determined, the CC energy is given by the expectation value of $\bar{H}$ with respect to the reference configuration,
\begin{equation}
    \label{EQN:CC_Engergy}
    E_{0} = \langle \Phi_0 | {\bar{H}} | \Phi_0 \rangle.
\end{equation}

Given optimal cluster amplitudes obtained from solving Eq.~\ref{EQN:CC_amplitudes}, excited-state information can be determined using the EOMCC approach, wherein the excited-state energies are given by the eigenvalues of the similarity-transformed Hamiltonian. One of the major strengths of the EOMCC formalism is that one has great flexibility in terms of the many-particle basis in which $\bar{H}$ is expanded, and different choices give access to different particle-number of spin-symmetry sectors of Fock space (see Ref.~\citenum{Krylov08_433} for a review on this topic). For example, energies and wave functions of the $(N-2)$-electron ({\em i.e.}, doubly ionized) states can be obtained from DIP-EOMCC, where the $K$-th doubly ionized state ($K>0$) is parametrized as 
\begin{equation}
\label{eqn:eom_wfn}
    \ket*{\Psi_K^{(N-2)}} = \hat{R}_K \ket*{\Psi_0^{(N)}} = \hat{R}_K \exp(\hat{T}) \ket{\Phi_0},
\end{equation}
In the DIP-EOMCC approach, the excitation operator $\hat{R}_K$ is taken to be a linear, non-particle-conserving operator that removes two electron from the $N$-electron state. We have
\begin{equation}
\label{eqn:eom_r_operator}
    \hat{R}_K = \sum_{n=2}^{ M^\prime} \hat{R}_{K,n},\;
    \hat{R}_{K,n} = \frac{1}{n!(n-2)!} r_{K,a_3 \ldots a_{n}}^{\phantom{K,}i_1 \ldots i_n}
                     \hat{a}_{i_2} \hat{a}_{i_1} \prod_{k=3}^{n} (\hat{a}^{a_k} \hat{a}_{i_k}),
\end{equation}
where { $M^\prime$ is a truncation level for the EOM operator, which is independent of the truncation choice in the underlying CC calculation.
Setting $M^\prime = M+1$}, Eq.~\ref{eqn:eom_r_operator} leads to the {``standard''} DIP-EOMCCSD {($M = 2$)} and DIP-EOMCCSDT {($M = 3$)} approaches, which include up to three-hole--one-particle ($3h1p$) or four-hole--two-particle ($4h2p$) transitions in $\hat{R}_K$, respectively.
{ In this work, we also examine the DIP-EOMCCSD($4h2p$) approach,\cite{Piecuch13_194102,Piecuch14_868,Piecuch21_e1966534} where we employ up to $4h2p$ excitations in the EOMCC calculation ($M^\prime = 4$) on top of CCSD similarity-transformed Hamiltonian ($M = 2$).}
Inserting Eq.~\ref{eqn:eom_wfn} into the Schrödinger equation gives
\begin{equation}
\label{eqn:eom_eigenvalue}
    \bar{H} \hat{R}_K \ket{\Phi_0} = E_K \hat{R}_K \ket{\Phi_0}
\end{equation}
which is a non-Hermitian eigenvalue problem that can be solved for the energies of the doubly ionized states, $E_K$. The double IP values are then given by $\omega_K = E_K - E_0$.

\section{Computational Details}

\label{SEC:COMPUTATIONAL_DETAILS}

Section \ref{SEC:RESULTS} begins with a direct comparison between the DIP values determined using mmfX2C-DIP-EOMCCSD and 4c DIP-EOMCCSD using various relativistic Hamiltonians. The 4c DIP-EOMCCSD data for this evaluation were taken from Ref.~\citenum{Pal20_104302}. For consistency, the present mmfX2C-based calculations use the same basis sets and truncation scheme as were used in Ref.~\citenum{Pal20_104302}. All basis sets are of at least triple-zeta quality, mostly in the Dyall family, with high-lying virtual spinors (with orbital energies greater than 500 $E_{\rm h}$) excluded from the correlated part of the calculation. All electrons were correlated in these calculations. The reader is referred to Table I of Ref. \citenum{Pal20_104302} for additional details.  Subsequent studies using mmfX2C- and 1eX2C-DIP-EOMCC examined a variety of basis sets, including the X2C-SVPall-2c and X2C-TZVPPall-2c,\cite{Weigend17_3696} Dyall.acv$n$z,\cite{Dyall02_335, Dyall06_441,Dyall16_128, Dyall12_1217, Dyall12_1172, DyallBasisZenodo} and ANO-RCC-V$n$ZP\cite{Widmark05_6575, roos04_2851} families, where $n$ is the cardinal number. Unlike above, these studies were carried out using the full set of virtual spinors. We also use the frozen-core approximation, where only electrons in the valence shell plus one inner shell are correlated.

The X2C-DIP-EOMCCSD calculations were carried out using the Cholesky decomposition (CD) approximation to the non-relativistic electron repulsion integrals (ERIs), with a decomposition threshold of  $1\times10^{-4}$ $E_{\rm h}$. It is well known that use of the CD approximation in ground-state CCSD calculations is an effective way to reduce memory requirements, for example, by constructing and storing only subblocks of the four-virtual-index part of the ERI tensor.\cite{Sherrill13_2687} On the other hand, the EOM part of DIP-EOMCCSD does not depend on any parts of the ERI tensor involving more than three virtual spinor labels, so there is less direct benefit to the use of the CD approximation in EOM part of the DIP-EOMCCSD algorithm. The only $ov^3$-sized ERI term (where $o$ and $v$ represent the number of occupied and virtual spinors, respectively) is used in the construction of smaller $\bar{H}$ intermediate quantities that enter into the iterative parts of the EOM algorithm, which itself only involves quantities with two or fewer virtual spinor labels. Calculations carried out at the X2C-DIP-EOMCCSDT level of theory did not make use of the CD approximation to the ERI tensor. 

All calculations reported in this work were performed with the DIP-EOMCC code implemented in a development branch of Chronus Quantum\cite{Li20_e1436} using the TiledArray tensor algebra framework\cite{TiledArray}. Equations and corresponding TiledArray expressions were generated using the \texttt{p$^\dagger$q} package,\cite{DePrince21_e1954709,DePrince25_2501.08882} and expressions for DIP-EOMCCSD $\bar{H}$ intermediates were modified to account for the use of the CD approximation. { Our DIP-EOMCC codes are tested numerically against the non-relativistic implementations in CCPy.\cite{CCpy,Piecuch25_061101} We also used CCPy to perform non-relativistic DIP-EOMCCSD and DIP-EOMCCSDT calculations without CD approximation, which we use to analyze basis set convergence behavior of these methods and to construct a composite scheme aimed at replicating experimental double IP data (see below).}

\section{Results and Discussion}

\label{SEC:RESULTS}

We begin with an assessment of the agreement between double IPs obtained from the present mmfX2C-DIP-EOMCCSD calculations and literature values computed using four-component (4c) DIP-EOM-CCSD.\cite{Pal20_104302} Table \ref{tab:DIP_dyall} contains vertical double IP values for noble gas atoms and diatomic molecules that were computed using these methods, with three different relativistic Hamiltonians. As mentioned above and discussed in Ref.~\citenum{Pal20_104302}, hydrogen atoms are described using the aug-cc-pVTZ basis set, while all other atoms are described using various Dyall-type basis sets of triple-zeta quality. In all cases, all electrons were correlated, but virtual molecular spinors with energies above 500 $E_{\rm h}$ were excluded from the correlated parts of the calculations. 
Using the Dirac--Coulomb Hamiltonian, we find excellent agreement between double IPs obtained from DC-X2C- and 4c-DIP-EOM-CCSD, with the largest discrepancies being only 0.003 eV in magnitude. As such, we are able to conclude the following. First, the error introduced via the CD approximation in mmfX2C-DIP-EOMCC is negligible. Second, because two-electron relativistic effects are only captured at the mean-field level in mmfX2C-DIP-EOMCC, correlation effects stemming from the relativistic two-electron part of the Hamiltonian must be small for these systems. Reference~\citenum{Pal20_104302} also considered 4c-DIP-EOMCCSD calculations on the krypton atom that included the Gaunt term in the Hamiltonian; we find excellent agreement between 4c- and DCG-X2C-DIP-EOMCCSD double IPs in this case. The changes in the double IPs induced by the Gaunt term are consistent between these methods to within 0.001 eV or less. {These results obtained using the DC and DCG Hamiltonians indicate that the mmfX2C scheme recovers highly accurate approximations to double IP values from 4c calculations, at least for the systems studied in this work. More general statements regarding the robustness of mmfX2C in this context will require additional direct comparisons between the 2c and 4c frameworks on other atomic and molecular systems. }
For krypton and the remaining systems, DCG-X2C-DIP-EOMCCSD calculations indicate that the Gaunt term always lowers the double IPs, by as little as -0.004 eV (for the $^1\Sigma^-$ state of Cl$_2$) or as much as -0.043 eV (for the $^1S_0$ state of xenon atom), with the average change being -0.017 eV. On the other hand, the gauge term in DCB-X2C-DIP-EOMCCSD universally raises the double IPs, by a fairly consistent amount for each species (0.001 - 0.005 eV). For a given atom or molecule, the largest deviations between the gaunt contributions to the double IPs for different states are no more than 0.001 eV. 

\begin{table}[]
\begin{minipage}{\columnwidth}
\caption{Double ionization potentials (eV) from X2C- and 4c-DIP-EOMCCSD calculations carried out using different 
relativistic Hamiltonians. Basis set and virtual space truncation information can be found in Ref.~\citenum{Pal20_104302}.}
    \centering
    \maybecompressarraystretch
    \begin{tabular*}{\columnwidth}{@{\extracolsep{\fill}}ccccccc}
    \hline\hline
    & \multirow{2}{*}{State} &\multicolumn{2}{c}{Dirac--Coulomb} &\multicolumn{2}{c}{+Gaunt}& 
    +gauge \\% & \multirow{2}{*}{experiment}\\
   \cline{3-4}\cline{5-6} \cline{7-7} 
    & &  X2C & 4c & X2C & 4c & X2C  \\    
   \hline
   \multirow{5}{*}{Ar} 
    &$^3P_2$ &43.448 & 43.448& -0.008&-& 0.002 \\%&43.389\\
    &$^3P_1$ &43.595 & 43.596& -0.014&-& 0.002 \\%&43.527\\
    &$^3P_0$ &43.656 & 43.657& -0.016&-& 0.002 \\%&43.584\\
    &$^1D_2$ &45.241 & 45.241& -0.012&-& 0.002 \\%&45.126\\
    &$^1S_0$ &47.695 & 47.694& -0.012&-& 0.002 \\%&47.514\\
    \hline
    \multirow{5}{*}{Kr} 
    &$^3P_2$ &38.342 & 38.341& -0.012& -0.012& 0.004 \\%&38.359\\
    &$^3P_1$ &38.930 & 38.930& -0.024& -0.024& 0.004 \\%&38.923\\
    &$^3P_0$ &39.027 & 39.028& -0.023& -0.023& 0.004 \\%&39.018\\
    &$^1D_2$ &40.218 & 40.218& -0.023& -0.022& 0.004 \\%&40.175\\
    &$^1S_0$ &42.566 & 42.566& -0.027& -0.027& 0.004 \\%&42.461\\
    \hline
    \multirow{5}{*}{Xe} 
    &$^3P_2$ &33.016 & 33.016 & -0.014&-& 0.004 \\%&33.105\\
    &$^3P_1$ &34.267 & 34.268 & -0.031&-& 0.004 \\%&34.319\\
    &$^3P_0$ &34.064 & 34.065 & -0.021&-& 0.005 \\%&34.113\\
    &$^1D_2$ &35.221 & 35.222 & -0.031&-& 0.005 \\%&35.225\\
    &$^1S_0$ &37.656 & 37.659 & -0.043&-& 0.005 \\%&37.581\\
    \hline      
    % \hline
    \multirow{4}{*}{Cl$_2$}
    & $^3\Sigma^-$ & 31.394&31.397& -0.007&-&0.002\\%&31.13\\
    & $^1\Delta$   & 31.905&31.907& -0.008&-&0.002\\%&31.74\\
    & $^1\Sigma^+$ & 32.292&32.294& -0.008&-&0.002\\%&32.12\\
    & $^1\Sigma^-$ & 33.322&33.319& -0.004&-&0.001\\%&32.97\\
    \hline
    \multirow{4}{*}{Br$_2$}
    & $^3\Sigma^-_{g0}$  &28.473&28.473& -0.011&-&0.003\\%&28.39\\
    & $^3\Sigma^-_{g1}$  &28.635 &- & -0.015 &- & 0.003 \\%&28.53\\
    & $^1\Delta_{g2}$  &29.041&29.041& -0.015&-&0.003\\%&28.91\\
    & $^1\Sigma^+_{g0}$  &29.519&29.519& -0.019&-&0.003\\%&29.38\\
    \hline
    \multirow{3}{*}{HBr}  
    & $^3\Sigma^-$ & 32.756 & 32.757 & -0.011 & -& 0.003\\%&32.62\\
    & $^1\Delta$   & 34.143 & 34.143 & -0.013 & -& 0.003\\%&33.95\\
    & $^1\Sigma^+$ & 35.429 & 35.429 & -0.015 & -& 0.003\\%&35.19\\
    \hline
    \multirow{5}{*}{HI}
    & $^3\Sigma_0^-$& 29.173 &29.174&-0.013 &-& 0.004\\%&29.15\\
    & $^1\Sigma_1^-$& 29.411 &29.412&-0.018 &-& 0.003\\%&29.37\\
    & $^1\Delta$    & 30.480 &30.481&-0.018 &-& 0.004\\%&30.39\\
    & $^1\Sigma^+$  & 31.800 &31.801&-0.022 &-& 0.004\\%&31.64\\
    \hline
    \hline
    \end{tabular*}
    \label{tab:DIP_dyall}
\end{minipage}
\end{table}

Given the excellent agreement between mmfX2C- and 4c-DIP-EOMCCSD, it is also interesting to consider the degree to which different X2C schemes ({\em i.e.}, 1eX2C versus mmfX2C) provide consistent double IP values. Table \ref{tab:1ex2c} provides the mean absolute deviation (MAD) between 1eX2C- and DCB-X2C-DIP-EOMCCSD derived double IPs for the same systems and states considered in Table \ref{tab:DIP_dyall}. Here, we use the row-dependent DCB-parameterized SNSO\cite{Li23_5785} scheme in the 1eX2C procedure. Unlike in Table \ref{tab:DIP_dyall}, calculations are carried out using ANO-RCC-$n$ZVP ($n$ = D, T, Q) basis sets, all virtual orbitals are correlated in the calculations, and we only correlate the electrons in the valence shell plus one inner shell. From these data, it appears that  double IP values obtained using different X2C schemes can differ by as much as 0.1 eV. These deviations decrease substantially (by more than a factor of three) when increasing the basis set from double-zeta to triple-zeta quality, but the deviations grow slightly when  increasing the basis set from triple-zeta to quadruple-zeta quality. This behavior is similar to the behavior observed in Ref.~\citenum{DePrince25_084110} for single IP values computed using 1eX2C- and DCB-X2C-IP-EOMCCSD. In that case, we observe that, for contracted basis sets such as the ANO-RCC family, 1eX2C- and DCB-X2C-IP-EOMCCSD derived IP values sometimes differed by as much as a few hundredths of 1 eV, but the agreement between these relativistic treatments generally improved with increased basis set size. In Ref.~\citenum{DePrince25_084110}, the discrepancy was traced to the recontraction of the basis after the X2C-HF procedure. Clearly, a similar discrepancy exists in the case of the double IPs considered here, although, the current problem seems a bit worse. Not only are the observed energy differences significantly larger than those observed in Ref.~\citenum{DePrince25_084110},  they are also much larger than the difference in double IPs from X2C- and 4c-DIP-EOMCCSD provided in Table \ref{tab:DIP_dyall}. Table \ref{tab:1ex2c} also provides mean absolute deviations in energy gaps (the difference between the lowest-energy double IP for a given atom or molecule and the higher-energy double IPs). 1eX2C- and DCB-X2C-DIP-EOMCCSD again provide results that differ on the order of 0.01 eV or more, suggesting that this basis set recontraction issue does not benefit much from a cancellation of errors when looking at energy differences. Therefore, we focus on the more complete mmfX2C framework for the remainder of this work.  

\begin{table}[]
\begin{minipage}{\columnwidth}
\caption{Mean absolute deviations (eV) between 1eX2C- and DCB-X2C-DIP-EOMCCSD derived double IP values and gaps between the lowest double IP value and the higher values.}
    \centering
    \begin{tabular*}{\columnwidth}{@{\extracolsep{\fill}}cccc}
    \hline\hline
    Assessed quantity & DZVP & TZVP & QZVP\\
    \hline
    Double IPs        & 0.104 & 0.031 & 0.034 \\ 
    Energy gaps       & 0.024 & 0.031 & 0.015 \\
    \hline 
    \hline
    \end{tabular*}
    \label{tab:1ex2c}
\end{minipage}
\end{table}

\begin{figure}[!htbp]
    \centering
    \includegraphics[width=\columnwidth]{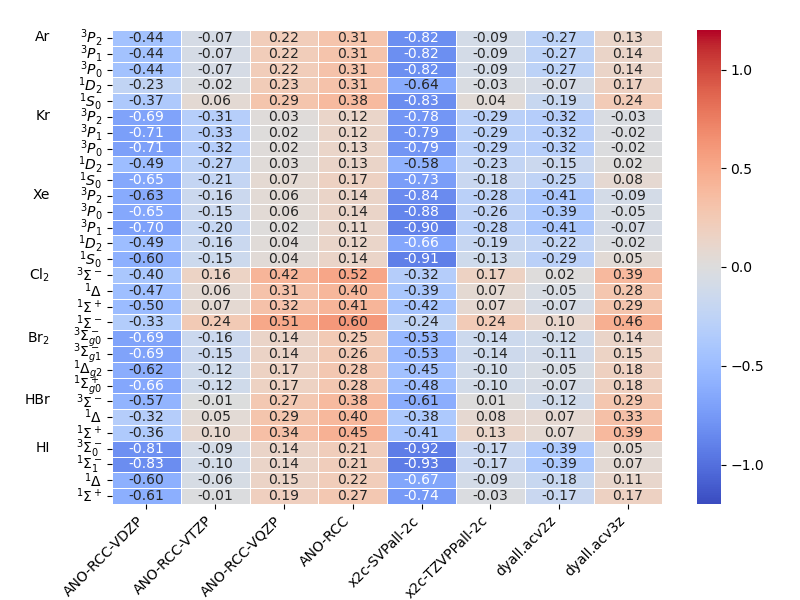}
    \caption{
        \label{fig:basis_set_effect}
        Errors in double ionization potentials (eV) calculated using ANO-RCC-V$n$ZP ($n$ = D, T, Q), X2C-$n$ZVPPall-2c ($n$ = D, T), and Dyall-acv$n$z ($n$ = D, T) basis sets. }
\end{figure}

We now assess the accuracy of double IPs from DCB-X2C-DIP-EOMCC calculations, as compared to experimentally determined values. Figure \ref{fig:basis_set_effect} provides errors in predicted double IP values obtained from DCB-X2C-DIP-EOMCCSD calculations carried out in various contracted (ANO-RCC-\cite{Widmark05_6575, roos04_2851} and x2c-type\cite{Weigend17_3696}) and uncontracted (Dyall-type\cite{Dyall02_335, Dyall06_441,Dyall16_128, Dyall12_1217, Dyall12_1172, DyallBasisZenodo}) basis sets. Results are provided for the same species and states tabulated in Table \ref{tab:DIP_dyall}, with experimental data taken from Refs.\citenum{NIST, Hall94_271, Eland08_270, Eland03_171, McNab04_179}. The raw computed and experimentally obtained double IP values can be found in the Supporting Information. 
Before analyzing the results of our DIP calculations, it is worth noting that previously reported DIP values~\cite{Pal20_104302,Piecuch25_061101} for the Cl$_2$, Br$_2$, HBr, and HI diatomics correspond to purely electronic vertical transitions, whereas the experimental double IPs reported in Refs.~\citenum{Hall94_271, Eland08_270, Eland03_171, McNab04_179} are extracted from vibrationally resolved threshold photoelectron spectra,
{ which account for the zero-point vibrational energy (ZPVE) differences between the neutral and dication states, and may involve adiabatic instead of direct vertical double ionizations.\cite{Hall94_271,McNab04_179,Eland08_270}}
The estimates for ZPVE effects, calculated as the differences between the harmonic vibrational frequencies of Cl$_2$, Br$_2$, HBr, and HI in their lowest dication and neutral states,\cite{NIST, Hall94_271, Eland08_270, Eland03_171, McNab04_179} amount to $-0.002$ (Br$_2$), $-0.027$ (Cl$_2$), $-0.047$ (HI), and $-0.123$ (HBr) eV. Accurate comparison against experimental data should take into account the geometrical relaxation and ZPVE effects, but in this work we report the purely electronic vertical DIP values consistent with past theoretical treatments.

As was the case for the data presented in Table \ref{tab:1ex2c}, the data in Fig.~\ref{fig:basis_set_effect} were generated by calculations that made use of the frozen core approximation ({\em i.e.}, we only correlate electrons in the valence shell plus one inner shell), with all virtual orbitals correlated. We begin by considering the convergence of the double IP values toward the complete basis set limit, with the truncated ANO-RCC-$n$ZVP ($n$ = D, T, Q) and full ANO-RCC basis sets. We find that the double IPs are severely underestimated within the double-zeta basis, with a mean absolute error (MAE) relative to experiment of more than 0.5 eV. Generally speaking, the double IP values increase with the size of the basis set, and DCB-X2C-DIP-EOMCCSD seems to agree best with experiment when combined with the triple-zeta-quality basis set (the MAE is 0.135 eV in this case). On the other hand,  increasing the basis set to the quadruple-zeta level, followed by the full ANO-RCC basis set leads to larger MAE values (0.175 eV and 0.269 eV, respectively). The fact that large-basis DCB-X2C-DIP-EOMCCSD overestimates the double IPs reflects the fact that DCB-X2C-DIP-EOMCCSD is not converged with respect to the correlation treatment.  We observe similar qualitative trends for DCB-X2C-DIP-EOMCCSD calculations carried out using other basis set families. Double IPs are underestimated using the smallest Dyall-type (dyall.acvdz) and x2c-type (x2c-SVPall-2c) basis sets (with MAEs of 0.204 eV and 0.661 eV, respectively), and the double IP values increase with increasing basis set size. At the triple-zeta level, the MAEs calculated using the dyall.acvtz and x2c-TZVPPall-2c basis sets fall to 0.159 eV and 0.149 eV, respectively.

\begin{figure}[!htbp]
    \centering
    \includegraphics[width=\columnwidth]{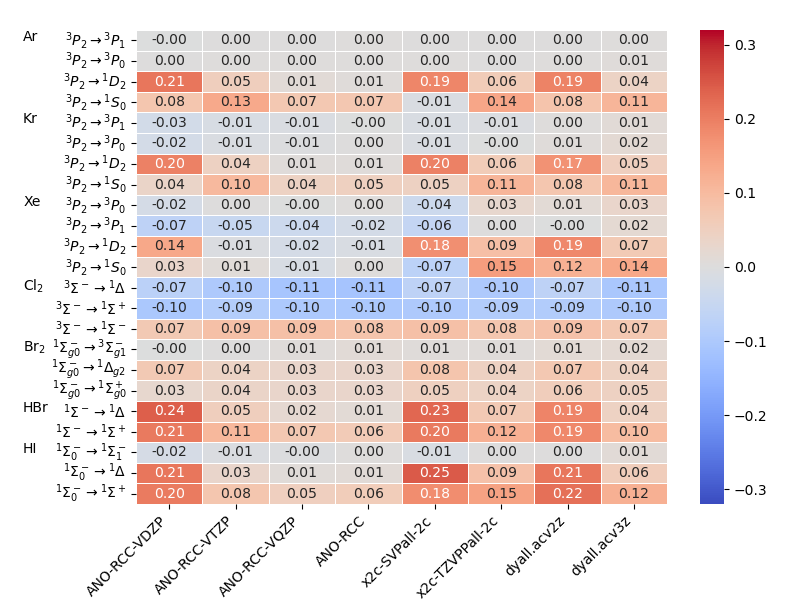}
    \caption{
        \label{fig:basis_set_effect_excitation}
        Errors in excitation energies (eV) for doubly ionized systems calculated using ANO-RCC-V$n$ZP ($n$ = D, T, Q), X2C-$n$ZVPPall-2c ($n$ = D, T), and Dyall-acv$n$z ($n$ = D, T) basis sets.}
\end{figure}

In addition to absolute double IP values, it is useful to consider the accuracy of DCB-X2C-DIP-EOMCCSD for predicting excitation energies in doubly ionized species, {\em i.e.}, the energy gaps between the lowest-energy state and the higher-energy states.  Figure \ref{fig:basis_set_effect_excitation} provides the energy gaps corresponding to the states considered in Fig.~\ref{fig:basis_set_effect}, using the same basis sets and correlation spaces. Unlike the absolute double IPs, we observe a clear trend indicating that the quality of the energy gaps improves with increasing basis set size. For example, proceeding through the ANO-RCC family, the MAEs in the energy gaps compared to experiment are 0.089 eV (ANO-RCC-DZVP), 0.046 eV (ANO-RCC-TZVP), 0.032 eV (ANO-RCC-QZVP), and 0.031 eV (full ANO-RCC). We see similar improvements in the MAEs when moving from double-zeta to triple-zeta quality sets using the x2c-type (0.090 eV to 0.062 eV) and Dyall-type (0.089 eV to 0.057 eV) basis sets. As stated above, the remaining error in the large-basis energy gaps is due to correlation effects that are missing at the DCB-X2C-DIP-EOMCCSD level of theory.

The data discussed thus far indicate that DCB-X2C-DIP-EOMCCSD calculations carried out in large  basis sets do not provide accurate predictions of experimentally-obtained double IP values for the systems considered in this work. In the full ANO-RCC basis set DCB-X2C-DIP-EOMCCSD overestimates the double IP values by 0.1--0.6 eV. Given the good agreement between mmfX2C- and 4c-DIP-EOMCCSD calculations in Table \ref{tab:DIP_dyall}, we conclude that the poor performance of DCB-X2C-DIP-EOMCCSD stems  not from the relativistic treatment but, rather, from a lack of higher-order correlation effects that are missing at this level of theory. To remedy the situation, we adopt the following composite protocol to  approximately account for higher-order correlation effects missing at the DCB-X2C-DIP-EOMCCSD level of theory. We define
\begin{equation}
\label{eqn:composite_dip}
    \mathrm{DIP}_K = \mathrm{DIP}_K^\mathrm{SD/full}
                    + \left[\mathrm{DIP}_K^\mathrm{SDT/\textcolor{black}{method}}
                    - \mathrm{DIP}_K^\mathrm{SD/\textcolor{black}{method}}\right]
\end{equation}
The first term on the right-hand side of Eq.~\ref{eqn:composite_dip} represents the double IP value obtained using DCB-X2C-DIP-EOMCCSD within the full ANO-RCC basis, and the subsequent terms { define a correction involving the difference between double DIP values obtained via SDT- and SD-level methods. We consider three such corrections based on: (i) DCB-X2C-DIP-EOMCCSDT and DCB-X2C-DIP-EOMCCSD in the ANO-RCC-VDZP basis set, (ii) DCB-X2C-DIP-EOMCCSDT and DCB-X2C-DIP-EOMCCSD in the ANO-RCC-VTZP basis set, and (iii) non-relativistic DIP-EOMCCSDT and DIP-EOMCCSD in either the full ANO-RCC basis set (for Ar and Kr) or the ANO-RCC-QZVP basis set (for Xe). }

Errors in double IP values for atomic systems that were computed using DCB-X2C-DIP-EOMCCSD / ANO-RCC and via Eq.~\ref{eqn:composite_dip} are tabulated in Table \ref{tab:4h2p}. { As we observed in Fig.~\ref{fig:basis_set_effect}, DCB-X2C-DIP-EOMCCSD consistently overestimates the double IP values in the full ANO-RCC basis set by 0.31--0.38 eV for Ar, 0.12--0.17 eV for Kr, and 0.11--0.14 eV for Xe. Higher-order correlation effects captured via DCB-X2C-DIP-EOMCCSDT carried out within the ANO-RCC-VDZP basis set [labeled as +SDT(DCB/DZ)] substantially reduce} the errors in the double IP values from DCB-X2C-DIP-EOMCCSD, sometimes by more than an order of magnitude. {For example, the mean absolute error in the double IP values from +SDT(DCB/DZ) is roughly 0.025 eV. However, the high accuracy we observe is due to a fortunate cancellation of errors, which is evident upon considering a similar correlation correction from DCB-X2C-DIP-EOMCCSDT carried out within the ANO-RCC-VTZP basis set [denoted as +SDT(DCB/TZ)]. For the argon and krypton atoms, we find that the double IP values from are consistently underestimated by the the +SDT(DCB/TZ) correction, by -0.068-- -0.135 eV and -0.083--0.159 eV for Ar and Kr, respectively. While our computational resource limitations prevented us from computing the +SDT(DCB/TZ) double IP value for the xenon atom, we expect similar errors of similar magnitude to arise in this case. Clearly, the correlation correction carried out in the double-zeta-quality basis is not converged with respect to the basis set size. At this point, it is worth noting that we also considered a correlation correction using the DCB-X2C-DIP-EOMCCSD($4h2p$) approach, which ignores correlation effects from $\hat{T}_3$, within the ANO-RCC-DZVP basis set. We found that double IP values from this approach are lower-bounds to those from +SDT(DCB/DZ), by roughly 0.01--0.04 eV, which can be rationalized as manifesting from an overstabilization of the doubly ionized state, relative to the neutral species. The double IP values from the DCB-X2C-DIP-EOMCCSD($4h2p$) based correlation correction can be found in the Supporting Information.}

{Lastly, we consider additional non-relativistic calculations at the DIP-EOMCCSD and DIP-EOMCCSDT levels in order to further investigate the convergence of the ANO-RCC-V$n$ZP ($n$ = D, T, Q) basis sets toward the full ANO-RCC limit. The following information can be gleaned from the resulting data, which are provided in the Supporting Information. First, energy differences between DIP-EOMCCSDT and DIP-EOMCCSD (i.e., a non-relativistic version of the correction term in Eq.~\ref{eqn:composite_dip}) computed using the ANO-RCC-VTZP basis set are converged to within 0.05 eV relative to the ANO-RCC-VQZP basis set in Ar, Kr, and Xe, and to within less than 0.1 eV relative to the full ANO-RCC set in the cases of Ar and Kr (calculations on Xe with the full ANO-RCC basis set were not possible with our resources). Second, when comparing to results from X2C calculations, the non-relativistic correlation correction is similar in magnitude to the relativistic ones within the ANO-RCC-VTZP basis set, at least for Ar and Kr where we could perform DCB-X2C-DIP-EOMCCSDT. Put together, these observations serve as a justification for using the difference between DIP-EOMCCSDT and DIP-EOMCCSD double IP values in the full ANO-RCC basis set in Eq.~\ref{eqn:composite_dip}; these data are under the +SDT(NR/full) heading in Table \ref{tab:4h2p}. Like the double IP values from +SDP(DCB/TZ), double IP values from the +SDT(NR/full) approach are consistently too low compared to experiment, this time by $-0.135$ -- $-0.182$ eV (Ar), $-0.176$ -- $-0.231$ eV (Kr), or $-0.270$ -- $-0.298$ eV (Xe). The $\approx$ $-0.1$ -- $-0.3$ eV error in the +SDT(DCB/TZ) and +SDT(NR/full) are likely due to the choice of basis set family (in this study, ANO-RCC type). Previous work using X2C-based CC\cite{DePrinceIII24_6521} and EOMCC\cite{DePrince25_084110} has indicated that the energy landscape in CC/EOMCC calculations that include spin--orbit coupling interactions can be significantly affected by the choice of basis set family and contraction scheme. That said, it is also worth mentioning that the 0.1--0.3 eV we observe here errors likely do not arise from basis set recontraction, as the largest recontraction errors observed in Ref.~\citenum{DePrince25_084110} were only on the order of 0.02 eV. The DCB-X2C-DIP-EOMCCSD and DCB-X2C-DIP-EOMCCSDT data, along with the non-relativistic DIP-EOMCCSD/EOMCCSDT data obtained using CCPy, are compiled in the Supporting Information. The Supporting Information also includes alternative composite approaches seeking to combine relativistic / non-relativistic DIP-EOMCCSD and DIP-EOMCCSDT data to approach the experimental double IP values.}

\begin{table*}[!htpb]
\begin{minipage}{\linewidth}
\caption{Experimentally-obtained double IP values (eV)\cite{NIST} and errors in double IP values (eV) calculated using DCB-X2C-DIP-EOMCCSD in the ANO-RCC basis set (labeled SD) and via Eq. ~\ref{eqn:composite_dip} (labeled {+}SDT). }
    \centering
    \begin{tabular*}{\linewidth}{@{\extracolsep{\fill}}ccccccc}
    \hline\hline
    & State & SD & {+SDT(DCB/DZ)} & {+SDT(DCB/TZ)} & {+SDT(NR/full)} & experiment\\
    \hline
   \multirow{5}{*}{Ar}
    &$^3P_2$ &  0.307 &  0.000  & {-0.133} & {-0.182} & 43.389  \\
    &$^3P_1$ &  0.309 &  -0.001 & {-0.135} & {-0.179} & 43.527  \\
    &$^3P_0$ &  0.311 &  0.000  & {-0.135} & {-0.178} & 43.584  \\
    &$^1D_2$ &  0.313 &  0.005  & {-0.107} & {-0.158} & 45.126  \\
    &$^1S_0$ &  0.379 &  0.090  & {-0.068} & {-0.135} & 47.514  \\
    \hline
    \multirow{5}{*}{Kr}  
    &$^3P_2$ &  0.125 &  -0.036 & {-0.143} & {-0.229} & 38.359  \\
    &$^3P_1$ &  0.124 &  -0.047 & {-0.159} & {-0.231} & 38.923  \\
    &$^3P_0$ &  0.127 &  -0.043 & {-0.154} & {-0.227} & 39.018  \\
    &$^1D_2$ &  0.133 &  -0.028 & {-0.123} & {-0.203} & 40.175  \\
    &$^1S_0$ &  0.173 &  0.043  & {-0.083} & {-0.176} & 42.461  \\
    \hline
    \multirow{5}{*}{Xe}
    &$^3P_2$ &  0.135 &   0.016 & {---\footnotemark[1]} & {-0.274\footnotemark[2]} & 33.105  \\
    &$^3P_{0}$ &  0.139 &   0.009 & {---\footnotemark[1]} & {-0.270\footnotemark[2]} & 34.113  \\
    &$^3P_{1}$ &  0.110 &  -0.018 & {---\footnotemark[1]} & {-0.298\footnotemark[2]} & 34.319  \\
    &$^1D_2$ &  0.123 &  -0.015 & {---\footnotemark[1]} & {-0.283\footnotemark[2]} & 35.225  \\
    &$^1S_0$ &  0.138 &  -0.016 & {---\footnotemark[1]} & {-0.294\footnotemark[2]} & 37.581  \\
    \hline 
    \hline
    \end{tabular*}
    \footnotetext[1]{DCB-EOMCCSDT/ANO-RCC-TZ calculations were not performed for Xe.}
    \footnotetext[2]{The correction was derived from non-relativitic calculations performed in the ANO-RCC-VQZP basis set.}
    \label{tab:4h2p}
\end{minipage}
\end{table*}

\section{Conclusions}

\label{SEC:CONCLUSIONS}

We have implemented relativistic formulations of DIP-EOMCCSD and DIP-EOMCCSDT within the 1eX2C and DC-, DCG-, and DCB-X2C frameworks. Direct comparisons against full 4c-DIP-EOMCCSD calculations show excellent agreement with DC(G)-X2C-DIP-EOMCCSD, suggesting, at least for the systems studied herein, two-electron relativistic effects are well-described by the mean-field treatment in mmfX2C, and remaining relativistic two-electron and electron-positron correlation effects are negligible. A subsequent basis set study on vertical double IPs for noble gas and diatomic species has shown that DCB-X2C-DIP-EOMCCSD tends to overestimate double IP values in the limit of a complete one-electron basis, by more than 0.25 eV, on average. For atomic systems, we were able to demonstrate that a composite scheme whereby the dominant correlation effects are captured by large-basis DCB-X2C-DIP-EOMCCSD and remaining high-order correlation effects are approximately modeled via small-basis DCB-X2C-DIP-EOMCCSDT brings the double IP values into excellent agreement with experiment; for Xe atom, for example, absolute errors in double IP values from this approach are less than 0.02 eV.
{ However, we found the ANO-RCC family of basis sets used in our composite approach to have poor convergence behavior in terms of DCB-X2C-DIP-EOMCC calculations, as the estimates computed using the non-relativistic DIP-EOMCC approach at the large basis set limit indicates a larger 0.1--0.2 eV error relative to experimental data.}

\vspace{0.5cm}

{\bf Supporting Information} Double ionization potentials from all-electron versus frozen core DCB-X2C-DIP-EOMCCSD calculations, as well as double ionization potentials and excitation energies of doubly ionized species computed using 1eX2C- and DCB-X2C-DIP-EOMCCSD within the frozen core approximation.

\vspace{0.5cm}

\begin{acknowledgments}This material is based upon work supported by the U.S. Department of Energy, Office of Science, Office of Advanced Scientific Computing Research and Office of Basic Energy Sciences, Scientific Discovery through the Advanced Computing (SciDAC) program under Award No. DE-SC0022263. The Chronus Quantum software infrastructure development is supported by the Office of Advanced Cyberinfrastructure, National Science Foundation (Grant Nos. OAC-2103717 and OAC-2103705). This project used resources of the National Energy Research Scientific Computing Center, a DOE Office of Science User Facility supported by the Office of Science of the U.S. DOE under Contract No. DE-AC02-05CH11231 using NERSC award ERCAP-0027762 and ERCAP-0032454.\\ 
\end{acknowledgments}

\noindent {\bf CONFLICT OF INTEREST}\\

     The authors have no conflicts to disclose.\\
     
\noindent {\bf DATA AVAILABILITY}\\

    The data that support the findings of this study are available from the corresponding author upon reasonable request.

\bibliography{main, deprince,rdm,combine}

%aipnum4-2.bst 2019-01-14 (MD) hand-edited version of apsrev4-1.bst
\begin{thebibliography}{111}%
\makeatletter
\providecommand \@ifxundefined [1]{%
 \@ifx{#1\undefined}
}%
\providecommand \@ifnum [1]{%
 \ifnum #1\expandafter \@firstoftwo
 \else \expandafter \@secondoftwo
 \fi
}%
\providecommand \@ifx [1]{%
 \ifx #1\expandafter \@firstoftwo
 \else \expandafter \@secondoftwo
 \fi
}%
\providecommand \natexlab [1]{#1}%
\providecommand \enquote  [1]{``#1''}%
\providecommand \bibnamefont  [1]{#1}%
\providecommand \bibfnamefont [1]{#1}%
\providecommand \citenamefont [1]{#1}%
\providecommand \href@noop [0]{\@secondoftwo}%
\providecommand \href [0]{\begingroup \@sanitize@url \@href}%
\providecommand \@href[1]{\@@startlink{#1}\@@href}%
\providecommand \@@href[1]{\endgroup#1\@@endlink}%
\providecommand \@sanitize@url [0]{\catcode `\\12\catcode `\$12\catcode
  `\&12\catcode `\#12\catcode `\^12\catcode `\_12\catcode `\%12\relax}%
\providecommand \@@startlink[1]{}%
\providecommand \@@endlink[0]{}%
\providecommand \url  [0]{\begingroup\@sanitize@url \@url }%
\providecommand \@url [1]{\endgroup\@href {#1}{\urlprefix }}%
\providecommand \urlprefix  [0]{URL }%
\providecommand \Eprint [0]{\href }%
\providecommand \doibase [0]{https://doi.org/}%
\providecommand \selectlanguage [0]{\@gobble}%
\providecommand \bibinfo  [0]{\@secondoftwo}%
\providecommand \bibfield  [0]{\@secondoftwo}%
\providecommand \translation [1]{[#1]}%
\providecommand \BibitemOpen [0]{}%
\providecommand \bibitemStop [0]{}%
\providecommand \bibitemNoStop [0]{.\EOS\space}%
\providecommand \EOS [0]{\spacefactor3000\relax}%
\providecommand \BibitemShut  [1]{\csname bibitem#1\endcsname}%
\let\auto@bib@innerbib\@empty
%</preamble>
\bibitem [{\citenamefont {Coester}(1958)}]{Coester58_421}%
  \BibitemOpen
  \bibfield  {author} {\bibinfo {author} {\bibfnamefont {F.}~\bibnamefont
  {Coester}},\ }\bibfield  {title} {\enquote {\bibinfo {title} {Bound states of
  a many-particle system},}\ }\href
  {https://doi.org/https://doi.org/10.1016/0029-5582(58)90280-3} {\bibfield
  {journal} {\bibinfo  {journal} {Nucl. Phys.}\ }\textbf {\bibinfo {volume}
  {7}},\ \bibinfo {pages} {421--424} (\bibinfo {year} {1958})}\BibitemShut
  {NoStop}%
\bibitem [{\citenamefont {Coester}\ and\ \citenamefont
  {K{\"u}mmel}(1960)}]{Kuemmel60_477}%
  \BibitemOpen
  \bibfield  {author} {\bibinfo {author} {\bibfnamefont {F.}~\bibnamefont
  {Coester}}\ and\ \bibinfo {author} {\bibfnamefont {H.}~\bibnamefont
  {K{\"u}mmel}},\ }\bibfield  {title} {\enquote {\bibinfo {title} {Short-range
  correlations in nuclear wave functions},}\ }\href
  {https://doi.org/https://doi.org/10.1016/0029-5582(60)90140-1} {\bibfield
  {journal} {\bibinfo  {journal} {Nucl. Phys.}\ }\textbf {\bibinfo {volume}
  {17}},\ \bibinfo {pages} {477--485} (\bibinfo {year} {1960})}\BibitemShut
  {NoStop}%
\bibitem [{\citenamefont {{\v{C}}{\'\i}{\v{z}}ek}(1966)}]{Cizek66_4256}%
  \BibitemOpen
  \bibfield  {author} {\bibinfo {author} {\bibfnamefont {J.}~\bibnamefont
  {{\v{C}}{\'\i}{\v{z}}ek}},\ }\bibfield  {title} {\enquote {\bibinfo {title}
  {On the correlation problem in atomic and molecular systems. calculation of
  wavefunction components in ursell‐type expansion using quantum‐field
  theoretical methods},}\ }\href {https://doi.org/10.1063/1.1727484} {\bibfield
   {journal} {\bibinfo  {journal} {J. Chem. Phys.}\ }\textbf {\bibinfo {volume}
  {45}},\ \bibinfo {pages} {4256--4266} (\bibinfo {year} {1966})}\BibitemShut
  {NoStop}%
\bibitem [{\citenamefont {{\v{C}}{\'\i}{\v{z}}ek}(1969)}]{Cizek69_35}%
  \BibitemOpen
  \bibfield  {author} {\bibinfo {author} {\bibfnamefont {J.}~\bibnamefont
  {{\v{C}}{\'\i}{\v{z}}ek}},\ }\bibfield  {title} {\enquote {\bibinfo {title}
  {On the use of the cluster expansion and the technique of diagrams in
  calculations of correlation effects in atoms and molecules},}\ }\href
  {https://doi.org/https://doi.org/10.1002/9780470143599.ch2} {\bibfield
  {journal} {\bibinfo  {journal} {Adv. Chem. Phys.}\ }\textbf {\bibinfo
  {volume} {14}},\ \bibinfo {pages} {35--89} (\bibinfo {year}
  {1969})}\BibitemShut {NoStop}%
\bibitem [{\citenamefont {Paldus}, \citenamefont {{\v{C}}{\'\i}{\v{z}}ek},\
  and\ \citenamefont {Shavitt}(1972)}]{Shavitt72_50}%
  \BibitemOpen
  \bibfield  {author} {\bibinfo {author} {\bibfnamefont {J.}~\bibnamefont
  {Paldus}}, \bibinfo {author} {\bibfnamefont {J.}~\bibnamefont
  {{\v{C}}{\'\i}{\v{z}}ek}},\ and\ \bibinfo {author} {\bibfnamefont
  {I.}~\bibnamefont {Shavitt}},\ }\bibfield  {title} {\enquote {\bibinfo
  {title} {Correlation problems in atomic and molecular systems. iv. extended
  coupled-pair many-electron theory and its application to the
  b${\mathrm{h}}_{3}$ molecule},}\ }\href
  {https://doi.org/10.1103/PhysRevA.5.50} {\bibfield  {journal} {\bibinfo
  {journal} {Phys. Rev. A}\ }\textbf {\bibinfo {volume} {5}},\ \bibinfo {pages}
  {50--67} (\bibinfo {year} {1972})}\BibitemShut {NoStop}%
\bibitem [{\citenamefont {Paldus}\ and\ \citenamefont {Li}(1999)}]{Li99_1}%
  \BibitemOpen
  \bibfield  {author} {\bibinfo {author} {\bibfnamefont {J.}~\bibnamefont
  {Paldus}}\ and\ \bibinfo {author} {\bibfnamefont {X.}~\bibnamefont {Li}},\
  }\bibfield  {title} {\enquote {\bibinfo {title} {A critical assessment of
  coupled cluster method in quantum chemistry},}\ }\href
  {https://doi.org/https://doi.org/10.1002/9780470141694.ch1} {\bibfield
  {journal} {\bibinfo  {journal} {Adv. Chem. Phys.}\ }\textbf {\bibinfo
  {volume} {110}},\ \bibinfo {pages} {1--175} (\bibinfo {year}
  {1999})}\BibitemShut {NoStop}%
\bibitem [{\citenamefont {Bartlett}\ and\ \citenamefont
  {Musia{\l}}(2007)}]{Musial07_291}%
  \BibitemOpen
  \bibfield  {author} {\bibinfo {author} {\bibfnamefont {R.~J.}\ \bibnamefont
  {Bartlett}}\ and\ \bibinfo {author} {\bibfnamefont {M.}~\bibnamefont
  {Musia{\l}}},\ }\bibfield  {title} {\enquote {\bibinfo {title}
  {Coupled-cluster theory in quantum chemistry},}\ }\href
  {https://doi.org/10.1103/RevModPhys.79.291} {\bibfield  {journal} {\bibinfo
  {journal} {Rev. Mod. Phys.}\ }\textbf {\bibinfo {volume} {79}},\ \bibinfo
  {pages} {291--352} (\bibinfo {year} {2007})}\BibitemShut {NoStop}%
\bibitem [{\citenamefont {Purvis}\ and\ \citenamefont
  {Bartlett}(1982)}]{Bartlett82_1910}%
  \BibitemOpen
  \bibfield  {author} {\bibinfo {author} {\bibfnamefont {G.~D.}\ \bibnamefont
  {Purvis}}\ and\ \bibinfo {author} {\bibfnamefont {R.~J.}\ \bibnamefont
  {Bartlett}},\ }\bibfield  {title} {\enquote {\bibinfo {title} {A full
  coupled-cluster singles and doubles model: The inclusion of disconnected
  triples},}\ }\href {https://doi.org/10.1063/1.443164} {\bibfield  {journal}
  {\bibinfo  {journal} {J. Chem. Phys.}\ }\textbf {\bibinfo {volume} {76}},\
  \bibinfo {pages} {1910--1918} (\bibinfo {year} {1982})}\BibitemShut {NoStop}%
\bibitem [{\citenamefont {Cullen}\ and\ \citenamefont
  {Zerner}(1982)}]{Zerner82_4088}%
  \BibitemOpen
  \bibfield  {author} {\bibinfo {author} {\bibfnamefont {J.~M.}\ \bibnamefont
  {Cullen}}\ and\ \bibinfo {author} {\bibfnamefont {M.~C.}\ \bibnamefont
  {Zerner}},\ }\bibfield  {title} {\enquote {\bibinfo {title} {{The linked
  singles and doubles model: An approximate theory of electron correlation
  based on the coupled‐cluster ansatz}},}\ }\href
  {https://doi.org/10.1063/1.444319} {\bibfield  {journal} {\bibinfo  {journal}
  {J. Chem. Phys.}\ }\textbf {\bibinfo {volume} {77}},\ \bibinfo {pages}
  {4088--4109} (\bibinfo {year} {1982})}\BibitemShut {NoStop}%
\bibitem [{\citenamefont {Noga}\ and\ \citenamefont
  {Bartlett}(1987)}]{Bartlett87_7041}%
  \BibitemOpen
  \bibfield  {author} {\bibinfo {author} {\bibfnamefont {J.}~\bibnamefont
  {Noga}}\ and\ \bibinfo {author} {\bibfnamefont {R.~J.}\ \bibnamefont
  {Bartlett}},\ }\bibfield  {title} {\enquote {\bibinfo {title} {{The full
  CCSDT model for molecular electronic structure}},}\ }\href
  {https://doi.org/10.1063/1.452353} {\bibfield  {journal} {\bibinfo  {journal}
  {J. Chem. Phys.}\ }\textbf {\bibinfo {volume} {86}},\ \bibinfo {pages}
  {7041--7050} (\bibinfo {year} {1987})},\ \bibinfo {note} {{\bf 89}, 3401
  (1988) [Erratum]}\BibitemShut {NoStop}%
\bibitem [{\citenamefont {Scuseria}\ and\ \citenamefont
  {Schaefer}(1988)}]{Schaefer88_382}%
  \BibitemOpen
  \bibfield  {author} {\bibinfo {author} {\bibfnamefont {G.~E.}\ \bibnamefont
  {Scuseria}}\ and\ \bibinfo {author} {\bibfnamefont {H.~F.}\ \bibnamefont
  {Schaefer}},\ }\bibfield  {title} {\enquote {\bibinfo {title} {A new
  implementation of the full ccsdt model for molecular electronic structure},}\
  }\href {https://doi.org/https://doi.org/10.1016/0009-2614(88)80110-6}
  {\bibfield  {journal} {\bibinfo  {journal} {Chem. Phys. Lett.}\ }\textbf
  {\bibinfo {volume} {152}},\ \bibinfo {pages} {382--386} (\bibinfo {year}
  {1988})}\BibitemShut {NoStop}%
\bibitem [{\citenamefont {Emrich}(1981)}]{Emrich81_379}%
  \BibitemOpen
  \bibfield  {author} {\bibinfo {author} {\bibfnamefont {K.}~\bibnamefont
  {Emrich}},\ }\bibfield  {title} {\enquote {\bibinfo {title} {An extension of
  the coupled cluster formalism to excited states ({{I}})},}\ }\href
  {https://doi.org/10.1016/0375-9474(81)90179-2} {\bibfield  {journal}
  {\bibinfo  {journal} {Nucl. Phys. A}\ }\textbf {\bibinfo {volume} {351}},\
  \bibinfo {pages} {379--396} (\bibinfo {year} {1981})}\BibitemShut {NoStop}%
\bibitem [{\citenamefont {Geertsen}, \citenamefont {Rittby},\ and\
  \citenamefont {Bartlett}(1989)}]{Bartlett89_57}%
  \BibitemOpen
  \bibfield  {author} {\bibinfo {author} {\bibfnamefont {J.}~\bibnamefont
  {Geertsen}}, \bibinfo {author} {\bibfnamefont {M.}~\bibnamefont {Rittby}},\
  and\ \bibinfo {author} {\bibfnamefont {R.~J.}\ \bibnamefont {Bartlett}},\
  }\bibfield  {title} {\enquote {\bibinfo {title} {The equation-of-motion
  coupled-cluster method: Excitation energies of be and co},}\ }\href
  {https://doi.org/https://doi.org/10.1016/0009-2614(89)85202-9} {\bibfield
  {journal} {\bibinfo  {journal} {Chem. Phys. Lett.}\ }\textbf {\bibinfo
  {volume} {164}},\ \bibinfo {pages} {57 -- 62} (\bibinfo {year}
  {1989})}\BibitemShut {NoStop}%
\bibitem [{\citenamefont {Stanton}\ and\ \citenamefont
  {Bartlett}(1993)}]{Bartlett93_7029}%
  \BibitemOpen
  \bibfield  {author} {\bibinfo {author} {\bibfnamefont {J.~F.}\ \bibnamefont
  {Stanton}}\ and\ \bibinfo {author} {\bibfnamefont {R.~J.}\ \bibnamefont
  {Bartlett}},\ }\bibfield  {title} {\enquote {\bibinfo {title} {The equation
  of motion coupled-cluster method. a systematic biorthogonal approach to
  molecular excitation energies, transition probabilities, and excited state
  properties},}\ }\href {https://doi.org/10.1063/1.464746} {\bibfield
  {journal} {\bibinfo  {journal} {J. Chem. Phys.}\ }\textbf {\bibinfo {volume}
  {98}},\ \bibinfo {pages} {7029--7039} (\bibinfo {year} {1993})}\BibitemShut
  {NoStop}%
\bibitem [{\citenamefont {Monkhorst}(1977)}]{Monkhorst77_421}%
  \BibitemOpen
  \bibfield  {author} {\bibinfo {author} {\bibfnamefont {H.~J.}\ \bibnamefont
  {Monkhorst}},\ }\bibfield  {title} {\enquote {\bibinfo {title} {{Calculation
  of properties with the coupled-cluster method}},}\ }\href
  {https://onlinelibrary.wiley.com/doi/abs/10.1002/qua.560120850} {\bibfield
  {journal} {\bibinfo  {journal} {Int. J. Quantum Chem.}\ }\textbf {\bibinfo
  {volume} {12}},\ \bibinfo {pages} {421--432} (\bibinfo {year}
  {1977})}\BibitemShut {NoStop}%
\bibitem [{\citenamefont {Dalgaard}\ and\ \citenamefont
  {Monkhorst}(1983)}]{Monkhorst83_1217}%
  \BibitemOpen
  \bibfield  {author} {\bibinfo {author} {\bibfnamefont {E.}~\bibnamefont
  {Dalgaard}}\ and\ \bibinfo {author} {\bibfnamefont {H.~J.}\ \bibnamefont
  {Monkhorst}},\ }\bibfield  {title} {\enquote {\bibinfo {title} {Some aspects
  of the time-dependent coupled-cluster approach to dynamic response
  functions},}\ }\href {https://doi.org/10.1103/PhysRevA.28.1217} {\bibfield
  {journal} {\bibinfo  {journal} {Phys. Rev. A}\ }\textbf {\bibinfo {volume}
  {28}},\ \bibinfo {pages} {1217--1222} (\bibinfo {year} {1983})}\BibitemShut
  {NoStop}%
\bibitem [{\citenamefont {Mukherjee}\ and\ \citenamefont
  {Mukherjee}(1979)}]{Mukherjee79_325}%
  \BibitemOpen
  \bibfield  {author} {\bibinfo {author} {\bibfnamefont {D.}~\bibnamefont
  {Mukherjee}}\ and\ \bibinfo {author} {\bibfnamefont {P.}~\bibnamefont
  {Mukherjee}},\ }\bibfield  {title} {\enquote {\bibinfo {title} {A
  response-function approach to the direct calculation of the transition-energy
  in a multiple-cluster expansion formalism},}\ }\href@noop {} {\bibfield
  {journal} {\bibinfo  {journal} {Chem. Phys.}\ }\textbf {\bibinfo {volume}
  {39}},\ \bibinfo {pages} {325--335} (\bibinfo {year} {1979})}\BibitemShut
  {NoStop}%
\bibitem [{\citenamefont {Sekino}\ and\ \citenamefont
  {Bartlett}(1984)}]{Bartlett84_255}%
  \BibitemOpen
  \bibfield  {author} {\bibinfo {author} {\bibfnamefont {H.}~\bibnamefont
  {Sekino}}\ and\ \bibinfo {author} {\bibfnamefont {R.~J.}\ \bibnamefont
  {Bartlett}},\ }\bibfield  {title} {\enquote {\bibinfo {title} {A linear
  response, coupled-cluster theory for excitation energy},}\ }\href
  {https://onlinelibrary.wiley.com/doi/abs/10.1002/qua.560260826} {\bibfield
  {journal} {\bibinfo  {journal} {Int. J. Quantum Chem.}\ }\textbf {\bibinfo
  {volume} {26}},\ \bibinfo {pages} {255--265} (\bibinfo {year}
  {1984})}\BibitemShut {NoStop}%
\bibitem [{\citenamefont {Takahashi}\ and\ \citenamefont
  {Paldus}(1986)}]{Paldus86_1486}%
  \BibitemOpen
  \bibfield  {author} {\bibinfo {author} {\bibfnamefont {M.}~\bibnamefont
  {Takahashi}}\ and\ \bibinfo {author} {\bibfnamefont {J.}~\bibnamefont
  {Paldus}},\ }\bibfield  {title} {\enquote {\bibinfo {title} {Time-dependent
  coupled cluster approach: {{Excitation}} energy calculation using an
  orthogonally spin-adapted formalism},}\ }\href
  {https://doi.org/10.1063/1.451241} {\bibfield  {journal} {\bibinfo  {journal}
  {J. Chem. Phys.}\ }\textbf {\bibinfo {volume} {85}},\ \bibinfo {pages}
  {1486--1501} (\bibinfo {year} {1986})}\BibitemShut {NoStop}%
\bibitem [{\citenamefont {Koch}\ and\ \citenamefont
  {J{\o}rgensen}(1990)}]{Jorgensen90_3333}%
  \BibitemOpen
  \bibfield  {author} {\bibinfo {author} {\bibfnamefont {H.}~\bibnamefont
  {Koch}}\ and\ \bibinfo {author} {\bibfnamefont {P.}~\bibnamefont
  {J{\o}rgensen}},\ }\bibfield  {title} {\enquote {\bibinfo {title} {Coupled
  cluster response functions},}\ }\href@noop {} {\bibfield  {journal} {\bibinfo
   {journal} {J. Chem. Phys.}\ }\textbf {\bibinfo {volume} {93}},\ \bibinfo
  {pages} {3333--3344} (\bibinfo {year} {1990})}\BibitemShut {NoStop}%
\bibitem [{\citenamefont {Koch}\ \emph {et~al.}(1990)\citenamefont {Koch},
  \citenamefont {Jensen}, \citenamefont {J{\o}rgensen},\ and\ \citenamefont
  {Helgaker}}]{Helgaker90_3345}%
  \BibitemOpen
  \bibfield  {author} {\bibinfo {author} {\bibfnamefont {H.}~\bibnamefont
  {Koch}}, \bibinfo {author} {\bibfnamefont {H.~J.~A.}\ \bibnamefont {Jensen}},
  \bibinfo {author} {\bibfnamefont {P.}~\bibnamefont {J{\o}rgensen}},\ and\
  \bibinfo {author} {\bibfnamefont {T.}~\bibnamefont {Helgaker}},\ }\bibfield
  {title} {\enquote {\bibinfo {title} {Excitation energies from the coupled
  cluster singles and doubles linear response function (ccsdlr). applications
  to be, ch+, co, and h2o},}\ }\href {https://doi.org/10.1063/1.458815}
  {\bibfield  {journal} {\bibinfo  {journal} {J. Chem. Phys.}\ }\textbf
  {\bibinfo {volume} {93}},\ \bibinfo {pages} {3345--3350} (\bibinfo {year}
  {1990})}\BibitemShut {NoStop}%
\bibitem [{\citenamefont {Nakatsuji}(1979)}]{Nakatsuji79_334}%
  \BibitemOpen
  \bibfield  {author} {\bibinfo {author} {\bibfnamefont {H.}~\bibnamefont
  {Nakatsuji}},\ }\bibfield  {title} {\enquote {\bibinfo {title} {Cluster
  expansion of the wavefunction. {{Calculation}} of electron correlations in
  ground and excited states by {{SAC}} and {{SAC CI}} theories},}\ }\href
  {https://www.sciencedirect.com/science/article/pii/0009261479851738}
  {\bibfield  {journal} {\bibinfo  {journal} {Chem. Phys. Lett.}\ }\textbf
  {\bibinfo {volume} {67}},\ \bibinfo {pages} {334--342} (\bibinfo {year}
  {1979})}\BibitemShut {NoStop}%
\bibitem [{\citenamefont {Piecuch}\ and\ \citenamefont
  {Kowalski}(2002)}]{Kowalski02_676}%
  \BibitemOpen
  \bibfield  {author} {\bibinfo {author} {\bibfnamefont {P.}~\bibnamefont
  {Piecuch}}\ and\ \bibinfo {author} {\bibfnamefont {K.}~\bibnamefont
  {Kowalski}},\ }\bibfield  {title} {\enquote {\bibinfo {title} {The
  {{State-Universal Multi-Reference Coupled-Cluster Theory}}: {{An Overview}}
  of {{Some Recent Advances}}},}\ }\href
  {https://www.mdpi.com/1422-0067/3/6/676} {\bibfield  {journal} {\bibinfo
  {journal} {Int. J. Mol. Sci.}\ }\textbf {\bibinfo {volume} {3}},\ \bibinfo
  {pages} {676--709} (\bibinfo {year} {2002})}\BibitemShut {NoStop}%
\bibitem [{\citenamefont {Lyakh}\ \emph {et~al.}(2012)\citenamefont {Lyakh},
  \citenamefont {Musia{\l}}, \citenamefont {Lotrich},\ and\ \citenamefont
  {Bartlett}}]{Bartlett12_182}%
  \BibitemOpen
  \bibfield  {author} {\bibinfo {author} {\bibfnamefont {D.~I.}\ \bibnamefont
  {Lyakh}}, \bibinfo {author} {\bibfnamefont {M.}~\bibnamefont {Musia{\l}}},
  \bibinfo {author} {\bibfnamefont {V.~F.}\ \bibnamefont {Lotrich}},\ and\
  \bibinfo {author} {\bibfnamefont {R.~J.}\ \bibnamefont {Bartlett}},\
  }\bibfield  {title} {\enquote {\bibinfo {title} {Multireference {{Nature}} of
  {{Chemistry}}: {{The Coupled-Cluster View}}},}\ }\href
  {https://doi.org/10.1021/cr2001417} {\bibfield  {journal} {\bibinfo
  {journal} {Chem. Rev.}\ }\textbf {\bibinfo {volume} {112}},\ \bibinfo {pages}
  {182--243} (\bibinfo {year} {2012})}\BibitemShut {NoStop}%
\bibitem [{\citenamefont {Evangelista}(2018)}]{Evangelista18_030901}%
  \BibitemOpen
  \bibfield  {author} {\bibinfo {author} {\bibfnamefont {F.~A.}\ \bibnamefont
  {Evangelista}},\ }\bibfield  {title} {\enquote {\bibinfo {title}
  {Perspective: {{Multireference}} coupled cluster theories of dynamical
  electron correlation},}\ }\href {https://doi.org/10.1063/1.5039496}
  {\bibfield  {journal} {\bibinfo  {journal} {J. Chem. Phys.}\ }\textbf
  {\bibinfo {volume} {149}},\ \bibinfo {pages} {030901} (\bibinfo {year}
  {2018})}\BibitemShut {NoStop}%
\bibitem [{\citenamefont {Casanova}\ \emph {et~al.}(2009)\citenamefont
  {Casanova}, \citenamefont {Slipchenko}, \citenamefont {Krylov},\ and\
  \citenamefont {{Head-Gordon}}}]{Head-Gordon09_044103}%
  \BibitemOpen
  \bibfield  {author} {\bibinfo {author} {\bibfnamefont {D.}~\bibnamefont
  {Casanova}}, \bibinfo {author} {\bibfnamefont {L.~V.}\ \bibnamefont
  {Slipchenko}}, \bibinfo {author} {\bibfnamefont {A.~I.}\ \bibnamefont
  {Krylov}},\ and\ \bibinfo {author} {\bibfnamefont {M.}~\bibnamefont
  {{Head-Gordon}}},\ }\bibfield  {title} {\enquote {\bibinfo {title} {Double
  spin-flip approach within equation-of-motion coupled cluster and
  configuration interaction formalisms: {{Theory}}, implementation, and
  examples},}\ }\href {https://doi.org/10.1063/1.3066652} {\bibfield  {journal}
  {\bibinfo  {journal} {J. Chem. Phys.}\ }\textbf {\bibinfo {volume} {130}},\
  \bibinfo {pages} {044103} (\bibinfo {year} {2009})}\BibitemShut {NoStop}%
\bibitem [{\citenamefont {Krylov}(2001)}]{Krylov01_375}%
  \BibitemOpen
  \bibfield  {author} {\bibinfo {author} {\bibfnamefont {A.~I.}\ \bibnamefont
  {Krylov}},\ }\bibfield  {title} {\enquote {\bibinfo {title} {Size-consistent
  wave functions for bond-breaking: The equation-of-motion spin-flip model},}\
  }\href {https://www.sciencedirect.com/science/article/pii/S0009261401002871}
  {\bibfield  {journal} {\bibinfo  {journal} {Chem. Phys. Lett.}\ }\textbf
  {\bibinfo {volume} {338}},\ \bibinfo {pages} {375--384} (\bibinfo {year}
  {2001})}\BibitemShut {NoStop}%
\bibitem [{\citenamefont {Krylov}(2006)}]{Krylov06_83}%
  \BibitemOpen
  \bibfield  {author} {\bibinfo {author} {\bibfnamefont {A.~I.}\ \bibnamefont
  {Krylov}},\ }\bibfield  {title} {\enquote {\bibinfo {title} {Spin-{{Flip
  Equation-of-Motion Coupled-Cluster Electronic Structure Method}} for a
  {{Description}} of {{Excited States}}, {{Bond Breaking}}, {{Diradicals}}, and
  {{Triradicals}}},}\ }\href {https://doi.org/10.1021/ar0402006} {\bibfield
  {journal} {\bibinfo  {journal} {Acc. Chem. Res.}\ }\textbf {\bibinfo {volume}
  {39}},\ \bibinfo {pages} {83--91} (\bibinfo {year} {2006})}\BibitemShut
  {NoStop}%
\bibitem [{\citenamefont {Krylov}(2008)}]{Krylov08_433}%
  \BibitemOpen
  \bibfield  {author} {\bibinfo {author} {\bibfnamefont {A.~I.}\ \bibnamefont
  {Krylov}},\ }\bibfield  {title} {\enquote {\bibinfo {title}
  {Equation-of-motion coupled-cluster methods for open-shell and electronically
  excited species: The hitchhiker's guide to fock space},}\ }\href
  {https://doi.org/10.1146/annurev.physchem.59.032607.093602} {\bibfield
  {journal} {\bibinfo  {journal} {Annu. Rev. Phys. Chem.}\ }\textbf {\bibinfo
  {volume} {59}},\ \bibinfo {pages} {433--462} (\bibinfo {year}
  {2008})}\BibitemShut {NoStop}%
\bibitem [{\citenamefont {Bartlett}\ and\ \citenamefont
  {Stanton}(1994)}]{Stanton94_65}%
  \BibitemOpen
  \bibfield  {author} {\bibinfo {author} {\bibfnamefont {R.~J.}\ \bibnamefont
  {Bartlett}}\ and\ \bibinfo {author} {\bibfnamefont {J.~F.}\ \bibnamefont
  {Stanton}},\ }\bibfield  {title} {\enquote {\bibinfo {title} {{Applications
  of Post-Hartree-Fock Methods: A Tutorial}},}\ }\href@noop {} {\ \textbf
  {\bibinfo {volume} {5}},\ \bibinfo {pages} {65--169} (\bibinfo {year}
  {1994})}\BibitemShut {NoStop}%
\bibitem [{\citenamefont {Nooijen}\ and\ \citenamefont
  {Snijders}(1992)}]{Snijders92_55}%
  \BibitemOpen
  \bibfield  {author} {\bibinfo {author} {\bibfnamefont {M.}~\bibnamefont
  {Nooijen}}\ and\ \bibinfo {author} {\bibfnamefont {J.~G.}\ \bibnamefont
  {Snijders}},\ }\bibfield  {title} {\enquote {\bibinfo {title} {Coupled
  cluster approach to the single-particle green's function},}\ }\href
  {https://doi.org/10.1002/qua.560440808} {\bibfield  {journal} {\bibinfo
  {journal} {Int. J. Quantum Chem.}\ }\textbf {\bibinfo {volume} {44}},\
  \bibinfo {pages} {55--83} (\bibinfo {year} {1992})}\BibitemShut {NoStop}%
\bibitem [{\citenamefont {Nooijen}\ and\ \citenamefont
  {Snijders}(1993)}]{Snijders93_15}%
  \BibitemOpen
  \bibfield  {author} {\bibinfo {author} {\bibfnamefont {M.}~\bibnamefont
  {Nooijen}}\ and\ \bibinfo {author} {\bibfnamefont {J.~G.}\ \bibnamefont
  {Snijders}},\ }\bibfield  {title} {\enquote {\bibinfo {title} {Coupled
  cluster green's function method: Working equations and applications},}\
  }\href {https://doi.org/10.1002/qua.560480103} {\bibfield  {journal}
  {\bibinfo  {journal} {Int. J. Quantum Chem.}\ }\textbf {\bibinfo {volume}
  {48}},\ \bibinfo {pages} {15--48} (\bibinfo {year} {1993})}\BibitemShut
  {NoStop}%
\bibitem [{\citenamefont {Stanton}\ and\ \citenamefont
  {Gauss}(1994)}]{Gauss94_8938}%
  \BibitemOpen
  \bibfield  {author} {\bibinfo {author} {\bibfnamefont {J.~F.}\ \bibnamefont
  {Stanton}}\ and\ \bibinfo {author} {\bibfnamefont {J.}~\bibnamefont
  {Gauss}},\ }\bibfield  {title} {\enquote {\bibinfo {title} {Analytic energy
  derivatives for ionized states described by the equation‐of‐motion
  coupled cluster method},}\ }\href {https://doi.org/10.1063/1.468022}
  {\bibfield  {journal} {\bibinfo  {journal} {J. Chem. Phys.}\ }\textbf
  {\bibinfo {volume} {101}},\ \bibinfo {pages} {8938--8944} (\bibinfo {year}
  {1994})}\BibitemShut {NoStop}%
\bibitem [{\citenamefont {Musia{\l}}, \citenamefont {Kucharski},\ and\
  \citenamefont {Bartlett}(2003)}]{Bartlett03_1128}%
  \BibitemOpen
  \bibfield  {author} {\bibinfo {author} {\bibfnamefont {M.}~\bibnamefont
  {Musia{\l}}}, \bibinfo {author} {\bibfnamefont {S.~A.}\ \bibnamefont
  {Kucharski}},\ and\ \bibinfo {author} {\bibfnamefont {R.~J.}\ \bibnamefont
  {Bartlett}},\ }\bibfield  {title} {\enquote {\bibinfo {title}
  {Equation-of-motion coupled cluster method with full inclusion of the
  connected triple excitations for ionized states: {{IP-EOM-CCSDT}}},}\ }\href
  {https://doi.org/10.1063/1.1527013} {\bibfield  {journal} {\bibinfo
  {journal} {J. Chem. Phys.}\ }\textbf {\bibinfo {volume} {118}},\ \bibinfo
  {pages} {1128--1136} (\bibinfo {year} {2003})}\BibitemShut {NoStop}%
\bibitem [{\citenamefont {Musia{\l}}\ and\ \citenamefont
  {Bartlett}(2004)}]{Bartlett04_210}%
  \BibitemOpen
  \bibfield  {author} {\bibinfo {author} {\bibfnamefont {M.}~\bibnamefont
  {Musia{\l}}}\ and\ \bibinfo {author} {\bibfnamefont {R.~J.}\ \bibnamefont
  {Bartlett}},\ }\bibfield  {title} {\enquote {\bibinfo {title} {{{EOM-CCSDT}}
  study of the low-lying ionization potentials of ethylene, acetylene and
  formaldehyde},}\ }\href {https://doi.org/10.1016/j.cplett.2003.11.059}
  {\bibfield  {journal} {\bibinfo  {journal} {Chem. Phys. Lett.}\ }\textbf
  {\bibinfo {volume} {384}},\ \bibinfo {pages} {210--214} (\bibinfo {year}
  {2004})}\BibitemShut {NoStop}%
\bibitem [{\citenamefont {Bomble}\ \emph {et~al.}(2005)\citenamefont {Bomble},
  \citenamefont {Saeh}, \citenamefont {Stanton}, \citenamefont {Szalay},
  \citenamefont {K{\'a}llay},\ and\ \citenamefont {Gauss}}]{Gauss05_154107}%
  \BibitemOpen
  \bibfield  {author} {\bibinfo {author} {\bibfnamefont {Y.~J.}\ \bibnamefont
  {Bomble}}, \bibinfo {author} {\bibfnamefont {J.~C.}\ \bibnamefont {Saeh}},
  \bibinfo {author} {\bibfnamefont {J.~F.}\ \bibnamefont {Stanton}}, \bibinfo
  {author} {\bibfnamefont {P.~G.}\ \bibnamefont {Szalay}}, \bibinfo {author}
  {\bibfnamefont {M.}~\bibnamefont {K{\'a}llay}},\ and\ \bibinfo {author}
  {\bibfnamefont {J.}~\bibnamefont {Gauss}},\ }\bibfield  {title} {\enquote
  {\bibinfo {title} {Equation-of-motion coupled-cluster methods for ionized
  states with an approximate treatment of triple excitations},}\ }\href
  {https://doi.org/10.1063/1.1884600} {\bibfield  {journal} {\bibinfo
  {journal} {J. Chem. Phys.}\ }\textbf {\bibinfo {volume} {122}},\ \bibinfo
  {pages} {154107} (\bibinfo {year} {2005})}\BibitemShut {NoStop}%
\bibitem [{\citenamefont {Gour}, \citenamefont {Piecuch},\ and\ \citenamefont
  {W{\l}och}(2005)}]{Wloch05_134113}%
  \BibitemOpen
  \bibfield  {author} {\bibinfo {author} {\bibfnamefont {J.~R.}\ \bibnamefont
  {Gour}}, \bibinfo {author} {\bibfnamefont {P.}~\bibnamefont {Piecuch}},\ and\
  \bibinfo {author} {\bibfnamefont {M.}~\bibnamefont {W{\l}och}},\ }\bibfield
  {title} {\enquote {\bibinfo {title} {Active-space equation-of-motion
  coupled-cluster methods for excited states of radicals and other open-shell
  systems: {{EA-EOMCCSDt}} and {{IP-EOMCCSDt}}},}\ }\href
  {https://doi.org/10.1063/1.2042452} {\bibfield  {journal} {\bibinfo
  {journal} {J. Chem. Phys.}\ }\textbf {\bibinfo {volume} {123}},\ \bibinfo
  {pages} {134113} (\bibinfo {year} {2005})}\BibitemShut {NoStop}%
\bibitem [{\citenamefont {Gour}, \citenamefont {Piecuch},\ and\ \citenamefont
  {W{\l}och}(2006)}]{Wloch06_2854}%
  \BibitemOpen
  \bibfield  {author} {\bibinfo {author} {\bibfnamefont {J.~R.}\ \bibnamefont
  {Gour}}, \bibinfo {author} {\bibfnamefont {P.}~\bibnamefont {Piecuch}},\ and\
  \bibinfo {author} {\bibfnamefont {M.}~\bibnamefont {W{\l}och}},\ }\bibfield
  {title} {\enquote {\bibinfo {title} {Extension of the active-space
  equation-of-motion coupled-cluster methods to radical systems: {{The
  EA-EOMCCSDt}} and {{IP-EOMCCSDt}} approaches},}\ }\href
  {https://doi.org/10.1002/qua.21112} {\bibfield  {journal} {\bibinfo
  {journal} {Int. J. Quantum Chem.}\ }\textbf {\bibinfo {volume} {106}},\
  \bibinfo {pages} {2854--2874} (\bibinfo {year} {2006})}\BibitemShut {NoStop}%
\bibitem [{\citenamefont {Gour}\ and\ \citenamefont
  {Piecuch}(2006)}]{Piecuch06_234107}%
  \BibitemOpen
  \bibfield  {author} {\bibinfo {author} {\bibfnamefont {J.~R.}\ \bibnamefont
  {Gour}}\ and\ \bibinfo {author} {\bibfnamefont {P.}~\bibnamefont {Piecuch}},\
  }\bibfield  {title} {\enquote {\bibinfo {title} {Efficient formulation and
  computer implementation of the active-space electron-attached and ionized
  equation-of-motion coupled-cluster methods},}\ }\href
  {https://doi.org/10.1063/1.2409289} {\bibfield  {journal} {\bibinfo
  {journal} {J. Chem. Phys.}\ }\textbf {\bibinfo {volume} {125}},\ \bibinfo
  {pages} {234107} (\bibinfo {year} {2006})}\BibitemShut {NoStop}%
\bibitem [{\citenamefont {Nooijen}\ and\ \citenamefont
  {Bartlett}(1995{\natexlab{a}})}]{Bartlett95_3629}%
  \BibitemOpen
  \bibfield  {author} {\bibinfo {author} {\bibfnamefont {M.}~\bibnamefont
  {Nooijen}}\ and\ \bibinfo {author} {\bibfnamefont {R.~J.}\ \bibnamefont
  {Bartlett}},\ }\bibfield  {title} {\enquote {\bibinfo {title} {Equation of
  motion coupled cluster method for electron attachment},}\ }\href
  {https://doi.org/10.1063/1.468592} {\bibfield  {journal} {\bibinfo  {journal}
  {J. Chem. Phys.}\ }\textbf {\bibinfo {volume} {102}},\ \bibinfo {pages}
  {3629--3647} (\bibinfo {year} {1995}{\natexlab{a}})}\BibitemShut {NoStop}%
\bibitem [{\citenamefont {Nooijen}\ and\ \citenamefont
  {Bartlett}(1995{\natexlab{b}})}]{Bartlett95_6735}%
  \BibitemOpen
  \bibfield  {author} {\bibinfo {author} {\bibfnamefont {M.}~\bibnamefont
  {Nooijen}}\ and\ \bibinfo {author} {\bibfnamefont {R.~J.}\ \bibnamefont
  {Bartlett}},\ }\bibfield  {title} {\enquote {\bibinfo {title} {Description of
  core‐excitation spectra by the open‐shell electron‐attachment
  equation‐of‐motion coupled cluster method},}\ }\href
  {https://doi.org/10.1063/1.469147} {\bibfield  {journal} {\bibinfo  {journal}
  {J. Chem. Phys.}\ }\textbf {\bibinfo {volume} {102}},\ \bibinfo {pages}
  {6735--6756} (\bibinfo {year} {1995}{\natexlab{b}})}\BibitemShut {NoStop}%
\bibitem [{\citenamefont {Musia{\l}}\ and\ \citenamefont
  {Bartlett}(2003)}]{Bartlett03_1901}%
  \BibitemOpen
  \bibfield  {author} {\bibinfo {author} {\bibfnamefont {M.}~\bibnamefont
  {Musia{\l}}}\ and\ \bibinfo {author} {\bibfnamefont {R.~J.}\ \bibnamefont
  {Bartlett}},\ }\bibfield  {title} {\enquote {\bibinfo {title}
  {Equation-of-motion coupled cluster method with full inclusion of connected
  triple excitations for electron-attached states: {{EA-EOM-CCSDT}}},}\ }\href
  {https://doi.org/10.1063/1.1584657} {\bibfield  {journal} {\bibinfo
  {journal} {J. Chem. Phys.}\ }\textbf {\bibinfo {volume} {119}},\ \bibinfo
  {pages} {1901--1908} (\bibinfo {year} {2003})}\BibitemShut {NoStop}%
\bibitem [{\citenamefont {Wladyslawski}\ and\ \citenamefont
  {Nooijen}(2002)}]{Nooijen02_65}%
  \BibitemOpen
  \bibfield  {author} {\bibinfo {author} {\bibfnamefont {M.}~\bibnamefont
  {Wladyslawski}}\ and\ \bibinfo {author} {\bibfnamefont {M.}~\bibnamefont
  {Nooijen}},\ }\bibfield  {title} {\enquote {\bibinfo {title} {The
  {{Photoelectron Spectrum}} of the {{NO3 Radical Revisited}}: {{A Theoretical
  Investigation}} of {{Potential Energy Surfaces}} and {{Conical
  Intersections}}},}\ }in\ \href {https://doi.org/10.1021/bk-2002-0828.ch004}
  {\emph {\bibinfo {booktitle} {Low-{{Lying Potential Energy Surfaces}}}}},\
  \bibinfo {series} {{{ACS Symposium Series}}}, Vol.\ \bibinfo {volume} {828}\
  (\bibinfo {year} {2002})\ Chap.~\bibinfo {chapter} {4}, pp.\ \bibinfo {pages}
  {65--92}\BibitemShut {NoStop}%
\bibitem [{\citenamefont {Nooijen}(2002)}]{Nooijen02_656}%
  \BibitemOpen
  \bibfield  {author} {\bibinfo {author} {\bibfnamefont {M.}~\bibnamefont
  {Nooijen}},\ }\bibfield  {title} {\enquote {\bibinfo {title} {State
  {{Selective Equation}} of {{Motion Coupled Cluster Theory}}: {{Some
  Preliminary Results}}},}\ }\href {https://www.mdpi.com/1422-0067/3/6/656}
  {\bibfield  {journal} {\bibinfo  {journal} {Int. J. Mol. Sci.}\ }\textbf
  {\bibinfo {volume} {3}},\ \bibinfo {pages} {656--675} (\bibinfo {year}
  {2002})}\BibitemShut {NoStop}%
\bibitem [{\citenamefont {Musia{\l}}, \citenamefont {Perera},\ and\
  \citenamefont {Bartlett}(2011)}]{Bartlett11_114108}%
  \BibitemOpen
  \bibfield  {author} {\bibinfo {author} {\bibfnamefont {M.}~\bibnamefont
  {Musia{\l}}}, \bibinfo {author} {\bibfnamefont {A.}~\bibnamefont {Perera}},\
  and\ \bibinfo {author} {\bibfnamefont {R.~J.}\ \bibnamefont {Bartlett}},\
  }\bibfield  {title} {\enquote {\bibinfo {title} {Multireference
  coupled-cluster theory: {{The}} easy way},}\ }\href
  {https://doi.org/10.1063/1.3567115} {\bibfield  {journal} {\bibinfo
  {journal} {J. Chem. Phys.}\ }\textbf {\bibinfo {volume} {134}},\ \bibinfo
  {pages} {114108} (\bibinfo {year} {2011})}\BibitemShut {NoStop}%
\bibitem [{\citenamefont {Ku{\'s}}\ and\ \citenamefont
  {Krylov}(2011)}]{Krylov11_084109}%
  \BibitemOpen
  \bibfield  {author} {\bibinfo {author} {\bibfnamefont {T.}~\bibnamefont
  {Ku{\'s}}}\ and\ \bibinfo {author} {\bibfnamefont {A.~I.}\ \bibnamefont
  {Krylov}},\ }\bibfield  {title} {\enquote {\bibinfo {title} {Using the
  charge-stabilization technique in the double ionization potential
  equation-of-motion calculations with dianion references},}\ }\href
  {https://doi.org/10.1063/1.3626149} {\bibfield  {journal} {\bibinfo
  {journal} {J. Chem. Phys.}\ }\textbf {\bibinfo {volume} {135}},\ \bibinfo
  {pages} {084109} (\bibinfo {year} {2011})}\BibitemShut {NoStop}%
\bibitem [{\citenamefont {Ku{\'s}}\ and\ \citenamefont
  {Krylov}(2012)}]{Krylov12_244109}%
  \BibitemOpen
  \bibfield  {author} {\bibinfo {author} {\bibfnamefont {T.}~\bibnamefont
  {Ku{\'s}}}\ and\ \bibinfo {author} {\bibfnamefont {A.~I.}\ \bibnamefont
  {Krylov}},\ }\bibfield  {title} {\enquote {\bibinfo {title} {De-perturbative
  corrections for charge-stabilized double ionization potential
  equation-of-motion coupled-cluster method},}\ }\href
  {https://doi.org/10.1063/1.4730296} {\bibfield  {journal} {\bibinfo
  {journal} {J. Chem. Phys.}\ }\textbf {\bibinfo {volume} {136}},\ \bibinfo
  {pages} {244109} (\bibinfo {year} {2012})}\BibitemShut {NoStop}%
\bibitem [{\citenamefont {Shen}\ and\ \citenamefont
  {Piecuch}(2013)}]{Piecuch13_194102}%
  \BibitemOpen
  \bibfield  {author} {\bibinfo {author} {\bibfnamefont {J.}~\bibnamefont
  {Shen}}\ and\ \bibinfo {author} {\bibfnamefont {P.}~\bibnamefont {Piecuch}},\
  }\bibfield  {title} {\enquote {\bibinfo {title} {{Doubly electron-attached
  and doubly ionized equation-of-motion coupled-cluster methods with
  4-particle–2-hole and 4-hole–2-particle excitations and their
  active-space extensions}},}\ }\href {https://doi.org/10.1063/1.4803883}
  {\bibfield  {journal} {\bibinfo  {journal} {J. Chem. Phys.}\ }\textbf
  {\bibinfo {volume} {138}},\ \bibinfo {pages} {194102} (\bibinfo {year}
  {2013})}\BibitemShut {NoStop}%
\bibitem [{\citenamefont {Shen}\ and\ \citenamefont
  {Piecuch}(2014)}]{Piecuch14_868}%
  \BibitemOpen
  \bibfield  {author} {\bibinfo {author} {\bibfnamefont {J.}~\bibnamefont
  {Shen}}\ and\ \bibinfo {author} {\bibfnamefont {P.}~\bibnamefont {Piecuch}},\
  }\bibfield  {title} {\enquote {\bibinfo {title} {Doubly electron-attached and
  doubly ionised equation-of-motion coupled-cluster methods with full and
  active-space treatments of 4-particle--2-hole and 4-hole--2-particle
  excitations: The role of orbital choices},}\ }\href
  {https://www.tandfonline.com/doi/abs/10.1080/00268976.2014.886397} {\bibfield
   {journal} {\bibinfo  {journal} {Mol. Phys.}\ }\textbf {\bibinfo {volume}
  {112}},\ \bibinfo {pages} {868--885} (\bibinfo {year} {2014})}\BibitemShut
  {NoStop}%
\bibitem [{\citenamefont {Gururangan}, \citenamefont {Dutta},\ and\
  \citenamefont {Piecuch}(2025)}]{Piecuch25_061101}%
  \BibitemOpen
  \bibfield  {author} {\bibinfo {author} {\bibfnamefont {K.}~\bibnamefont
  {Gururangan}}, \bibinfo {author} {\bibfnamefont {A.~K.}\ \bibnamefont
  {Dutta}},\ and\ \bibinfo {author} {\bibfnamefont {P.}~\bibnamefont
  {Piecuch}},\ }\bibfield  {title} {\enquote {\bibinfo {title} {Double
  ionization potential equation-of-motion coupled-cluster approach with full
  inclusion of 4-hole--2-particle excitations and three-body clusters},}\
  }\href {https://doi.org/10.1063/5.0253059} {\bibfield  {journal} {\bibinfo
  {journal} {J. Chem. Phys.}\ }\textbf {\bibinfo {volume} {162}},\ \bibinfo
  {pages} {061101} (\bibinfo {year} {2025})}\BibitemShut {NoStop}%
\bibitem [{\citenamefont {Dyall}(1997)}]{Dyall97_9618}%
  \BibitemOpen
  \bibfield  {author} {\bibinfo {author} {\bibfnamefont {K.~G.}\ \bibnamefont
  {Dyall}},\ }\bibfield  {title} {\enquote {\bibinfo {title} {Interfacing
  relativistic and nonrelativistic methods. {{I}}. {{Normalized}} elimination
  of the small component in the modified {{Dirac}} equation},}\ }\href
  {https://doi.org/10.1063/1.473860} {\bibfield  {journal} {\bibinfo  {journal}
  {J. Chem. Phys.}\ }\textbf {\bibinfo {volume} {106}},\ \bibinfo {pages}
  {9618--9626} (\bibinfo {year} {1997})}\BibitemShut {NoStop}%
\bibitem [{\citenamefont {Dyall}(1998)}]{Dyall98_4201}%
  \BibitemOpen
  \bibfield  {author} {\bibinfo {author} {\bibfnamefont {K.~G.}\ \bibnamefont
  {Dyall}},\ }\bibfield  {title} {\enquote {\bibinfo {title} {Interfacing
  relativistic and nonrelativistic methods. {II.} investigation of a low-order
  approximation},}\ }\href {https://doi.org/https://doi.org/10.1063/1.477026}
  {\bibfield  {journal} {\bibinfo  {journal} {J. Chem. Phys.}\ }\textbf
  {\bibinfo {volume} {109}},\ \bibinfo {pages} {4201--4208} (\bibinfo {year}
  {1998})}\BibitemShut {NoStop}%
\bibitem [{\citenamefont {Dyall}\ and\ \citenamefont
  {Enevoldsen}(1999)}]{Enevoldsen99_10000}%
  \BibitemOpen
  \bibfield  {author} {\bibinfo {author} {\bibfnamefont {K.~G.}\ \bibnamefont
  {Dyall}}\ and\ \bibinfo {author} {\bibfnamefont {T.}~\bibnamefont
  {Enevoldsen}},\ }\bibfield  {title} {\enquote {\bibinfo {title} {Interfacing
  relativistic and nonrelativistic methods. {III.} atomic 4-spinor expansions
  and integral approximations},}\ }\href
  {https://doi.org/https://doi.org/10.1063/1.480353} {\bibfield  {journal}
  {\bibinfo  {journal} {J. Chem. Phys.}\ }\textbf {\bibinfo {volume} {111}},\
  \bibinfo {pages} {10000--10007} (\bibinfo {year} {1999})}\BibitemShut
  {NoStop}%
\bibitem [{\citenamefont {Dyall}(2001)}]{Dyall01_9136}%
  \BibitemOpen
  \bibfield  {author} {\bibinfo {author} {\bibfnamefont {K.~G.}\ \bibnamefont
  {Dyall}},\ }\bibfield  {title} {\enquote {\bibinfo {title} {Interfacing
  relativistic and nonrelativistic methods. {IV.} {O}ne- and two-electron
  scalar approximations},}\ }\href
  {https://doi.org/https://doi.org/10.1063/1.1413512} {\bibfield  {journal}
  {\bibinfo  {journal} {J. Chem. Phys.}\ }\textbf {\bibinfo {volume} {115}},\
  \bibinfo {pages} {9136--9143} (\bibinfo {year} {2001})}\BibitemShut {NoStop}%
\bibitem [{\citenamefont {Filatov}\ and\ \citenamefont
  {Cremer}(2002)}]{Cremer02_259}%
  \BibitemOpen
  \bibfield  {author} {\bibinfo {author} {\bibfnamefont {M.}~\bibnamefont
  {Filatov}}\ and\ \bibinfo {author} {\bibfnamefont {D.}~\bibnamefont
  {Cremer}},\ }\bibfield  {title} {\enquote {\bibinfo {title} {A new
  quasi-relativistic approach for density functional theory based on the
  normalized elimination of the small component},}\ }\href
  {https://doi.org/https://doi.org/10.1016/S0009-2614(01)01357-4} {\bibfield
  {journal} {\bibinfo  {journal} {Chem. Phys. Lett.}\ }\textbf {\bibinfo
  {volume} {351}},\ \bibinfo {pages} {259--266} (\bibinfo {year}
  {2002})}\BibitemShut {NoStop}%
\bibitem [{\citenamefont {Kutzelnigg}\ and\ \citenamefont
  {Liu}(2005)}]{Liu05_241102}%
  \BibitemOpen
  \bibfield  {author} {\bibinfo {author} {\bibfnamefont {W.}~\bibnamefont
  {Kutzelnigg}}\ and\ \bibinfo {author} {\bibfnamefont {W.}~\bibnamefont
  {Liu}},\ }\bibfield  {title} {\enquote {\bibinfo {title} {Quasirelativistic
  theory equivalent to fully relativistic theory},}\ }\href
  {https://doi.org/10.1063/1.2137315} {\bibfield  {journal} {\bibinfo
  {journal} {J. Chem. Phys.}\ }\textbf {\bibinfo {volume} {123}},\ \bibinfo
  {pages} {241102} (\bibinfo {year} {2005})}\BibitemShut {NoStop}%
\bibitem [{\citenamefont {Liu}\ and\ \citenamefont
  {Peng}(2006)}]{Peng06_044102}%
  \BibitemOpen
  \bibfield  {author} {\bibinfo {author} {\bibfnamefont {W.}~\bibnamefont
  {Liu}}\ and\ \bibinfo {author} {\bibfnamefont {D.}~\bibnamefont {Peng}},\
  }\bibfield  {title} {\enquote {\bibinfo {title} {Infinite-order
  quasirelativistic density functional method based on the exact matrix
  quasirelativistic theory},}\ }\href
  {https://doi.org/https://doi.org/10.1063/1.2222365} {\bibfield  {journal}
  {\bibinfo  {journal} {J. Chem. Phys.}\ }\textbf {\bibinfo {volume} {125}},\
  \bibinfo {pages} {044102} (\bibinfo {year} {2006})}\BibitemShut {NoStop}%
\bibitem [{\citenamefont {Peng}\ \emph {et~al.}(2007)\citenamefont {Peng},
  \citenamefont {Liu}, \citenamefont {Xiao},\ and\ \citenamefont
  {Cheng}}]{Cheng07_104106}%
  \BibitemOpen
  \bibfield  {author} {\bibinfo {author} {\bibfnamefont {D.}~\bibnamefont
  {Peng}}, \bibinfo {author} {\bibfnamefont {W.}~\bibnamefont {Liu}}, \bibinfo
  {author} {\bibfnamefont {Y.}~\bibnamefont {Xiao}},\ and\ \bibinfo {author}
  {\bibfnamefont {L.}~\bibnamefont {Cheng}},\ }\bibfield  {title} {\enquote
  {\bibinfo {title} {Making four- and two-component relativistic density
  functional methods fully equivalent based on the idea of from atoms to
  molecule},}\ }\href {https://doi.org/https://doi.org/10.1063/1.2772856}
  {\bibfield  {journal} {\bibinfo  {journal} {J. Chem. Phys.}\ }\textbf
  {\bibinfo {volume} {127}},\ \bibinfo {pages} {104106} (\bibinfo {year}
  {2007})}\BibitemShut {NoStop}%
\bibitem [{\citenamefont {Ilia{\v{s}}}\ and\ \citenamefont
  {Saue}(2007)}]{Saue07_064102}%
  \BibitemOpen
  \bibfield  {author} {\bibinfo {author} {\bibfnamefont {M.}~\bibnamefont
  {Ilia{\v{s}}}}\ and\ \bibinfo {author} {\bibfnamefont {T.}~\bibnamefont
  {Saue}},\ }\bibfield  {title} {\enquote {\bibinfo {title} {An infinite-order
  relativistic hamiltonian by a simple one-step transformation},}\ }\href
  {https://doi.org/https://doi.org/10.1063/1.2436882} {\bibfield  {journal}
  {\bibinfo  {journal} {J. Chem. Phys.}\ }\textbf {\bibinfo {volume} {126}},\
  \bibinfo {pages} {064102} (\bibinfo {year} {2007})}\BibitemShut {NoStop}%
\bibitem [{\citenamefont {Liu}\ and\ \citenamefont
  {Peng}(2009)}]{Peng09_031104}%
  \BibitemOpen
  \bibfield  {author} {\bibinfo {author} {\bibfnamefont {W.}~\bibnamefont
  {Liu}}\ and\ \bibinfo {author} {\bibfnamefont {D.}~\bibnamefont {Peng}},\
  }\bibfield  {title} {\enquote {\bibinfo {title} {Exact two-component
  hamiltonians revisited},}\ }\href
  {https://doi.org/https://doi.org/10.1063/1.3159445} {\bibfield  {journal}
  {\bibinfo  {journal} {J. Chem. Phys.}\ }\textbf {\bibinfo {volume} {131}},\
  \bibinfo {pages} {031104} (\bibinfo {year} {2009})}\BibitemShut {NoStop}%
\bibitem [{\citenamefont {Sikkema}\ \emph {et~al.}(2009)\citenamefont
  {Sikkema}, \citenamefont {Visscher}, \citenamefont {Saue},\ and\
  \citenamefont {Ilia{\v{s}}}}]{Ilias09_124116}%
  \BibitemOpen
  \bibfield  {author} {\bibinfo {author} {\bibfnamefont {J.}~\bibnamefont
  {Sikkema}}, \bibinfo {author} {\bibfnamefont {L.}~\bibnamefont {Visscher}},
  \bibinfo {author} {\bibfnamefont {T.}~\bibnamefont {Saue}},\ and\ \bibinfo
  {author} {\bibfnamefont {M.}~\bibnamefont {Ilia{\v{s}}}},\ }\bibfield
  {title} {\enquote {\bibinfo {title} {The molecular mean-field approach for
  correlated relativistic calculations},}\ }\href
  {https://doi.org/https://doi.org/10.1063/1.3239505} {\bibfield  {journal}
  {\bibinfo  {journal} {J. Chem. Phys.}\ }\textbf {\bibinfo {volume} {131}},\
  \bibinfo {pages} {124116} (\bibinfo {year} {2009})}\BibitemShut {NoStop}%
\bibitem [{\citenamefont {Liu}(2010)}]{Liu10_1679}%
  \BibitemOpen
  \bibfield  {author} {\bibinfo {author} {\bibfnamefont {W.}~\bibnamefont
  {Liu}},\ }\bibfield  {title} {\enquote {\bibinfo {title} {Ideas of
  relativistic quantum chemistry},}\ }\href
  {https://doi.org/https://doi.org/10.1080/00268971003781571} {\bibfield
  {journal} {\bibinfo  {journal} {Mol. Phys.}\ }\textbf {\bibinfo {volume}
  {108}},\ \bibinfo {pages} {1679--1706} (\bibinfo {year} {2010})}\BibitemShut
  {NoStop}%
\bibitem [{\citenamefont {Li}, \citenamefont {Xiao},\ and\ \citenamefont
  {Liu}(2012)}]{Liu12_154114}%
  \BibitemOpen
  \bibfield  {author} {\bibinfo {author} {\bibfnamefont {Z.}~\bibnamefont
  {Li}}, \bibinfo {author} {\bibfnamefont {Y.}~\bibnamefont {Xiao}},\ and\
  \bibinfo {author} {\bibfnamefont {W.}~\bibnamefont {Liu}},\ }\bibfield
  {title} {\enquote {\bibinfo {title} {On the spin separation of algebraic
  two-component relativistic hamiltonians},}\ }\href
  {https://doi.org/https://doi.org/10.1063/1.4758987} {\bibfield  {journal}
  {\bibinfo  {journal} {J. Chem. Phys.}\ }\textbf {\bibinfo {volume} {137}},\
  \bibinfo {pages} {154114} (\bibinfo {year} {2012})}\BibitemShut {NoStop}%
\bibitem [{\citenamefont {Peng}\ \emph {et~al.}(2013)\citenamefont {Peng},
  \citenamefont {Middendorf}, \citenamefont {Weigend},\ and\ \citenamefont
  {Reiher}}]{Reiher13_184105}%
  \BibitemOpen
  \bibfield  {author} {\bibinfo {author} {\bibfnamefont {D.}~\bibnamefont
  {Peng}}, \bibinfo {author} {\bibfnamefont {N.}~\bibnamefont {Middendorf}},
  \bibinfo {author} {\bibfnamefont {F.}~\bibnamefont {Weigend}},\ and\ \bibinfo
  {author} {\bibfnamefont {M.}~\bibnamefont {Reiher}},\ }\bibfield  {title}
  {\enquote {\bibinfo {title} {An efficient implementation of two-component
  relativistic exact-decoupling methods for large molecules},}\ }\href
  {https://doi.org/https://doi.org/10.1063/1.4803693} {\bibfield  {journal}
  {\bibinfo  {journal} {J. Chem. Phys.}\ }\textbf {\bibinfo {volume} {138}},\
  \bibinfo {pages} {184105} (\bibinfo {year} {2013})}\BibitemShut {NoStop}%
\bibitem [{\citenamefont {Tecmer}\ \emph {et~al.}(2014)\citenamefont {Tecmer},
  \citenamefont {Gomes}, \citenamefont {Knecht},\ and\ \citenamefont
  {Visscher}}]{Visscher14_041107}%
  \BibitemOpen
  \bibfield  {author} {\bibinfo {author} {\bibfnamefont {P.}~\bibnamefont
  {Tecmer}}, \bibinfo {author} {\bibfnamefont {A.~S.~P.}\ \bibnamefont
  {Gomes}}, \bibinfo {author} {\bibfnamefont {S.}~\bibnamefont {Knecht}},\ and\
  \bibinfo {author} {\bibfnamefont {L.}~\bibnamefont {Visscher}},\ }\bibfield
  {title} {\enquote {\bibinfo {title} {Communication: Relativistic fock-space
  coupled cluster study of small building blocks of larger uranium
  complexes},}\ }\href {https://doi.org/https://doi.org/10.1063/1.4891801}
  {\bibfield  {journal} {\bibinfo  {journal} {J. Chem. Phys.}\ }\textbf
  {\bibinfo {volume} {141}},\ \bibinfo {pages} {041107} (\bibinfo {year}
  {2014})}\BibitemShut {NoStop}%
\bibitem [{\citenamefont {Egidi}\ \emph {et~al.}(2016)\citenamefont {Egidi},
  \citenamefont {Goings}, \citenamefont {Frisch},\ and\ \citenamefont
  {Li}}]{Li16_3711}%
  \BibitemOpen
  \bibfield  {author} {\bibinfo {author} {\bibfnamefont {F.}~\bibnamefont
  {Egidi}}, \bibinfo {author} {\bibfnamefont {J.~J.}\ \bibnamefont {Goings}},
  \bibinfo {author} {\bibfnamefont {M.~J.}\ \bibnamefont {Frisch}},\ and\
  \bibinfo {author} {\bibfnamefont {X.}~\bibnamefont {Li}},\ }\bibfield
  {title} {\enquote {\bibinfo {title} {Direct {{Atomic-Orbital-Based
  Relativistic Two-Component Linear Response Method}} for {{Calculating
  Excited-State Fine Structures}}},}\ }\href
  {https://doi.org/10.1021/acs.jctc.6b00474} {\bibfield  {journal} {\bibinfo
  {journal} {J. Chem. Theory Comput.}\ }\textbf {\bibinfo {volume} {12}},\
  \bibinfo {pages} {3711--3718} (\bibinfo {year} {2016})}\BibitemShut {NoStop}%
\bibitem [{\citenamefont {Goings}\ \emph {et~al.}(2016)\citenamefont {Goings},
  \citenamefont {Kasper}, \citenamefont {Egidi}, \citenamefont {Sun},\ and\
  \citenamefont {Li}}]{Li16_104107}%
  \BibitemOpen
  \bibfield  {author} {\bibinfo {author} {\bibfnamefont {J.~J.}\ \bibnamefont
  {Goings}}, \bibinfo {author} {\bibfnamefont {J.~M.}\ \bibnamefont {Kasper}},
  \bibinfo {author} {\bibfnamefont {F.}~\bibnamefont {Egidi}}, \bibinfo
  {author} {\bibfnamefont {S.}~\bibnamefont {Sun}},\ and\ \bibinfo {author}
  {\bibfnamefont {X.}~\bibnamefont {Li}},\ }\bibfield  {title} {\enquote
  {\bibinfo {title} {Real time propagation of the exact two component
  time-dependent density functional theory},}\ }\href
  {https://doi.org/10.1063/1.4962422} {\bibfield  {journal} {\bibinfo
  {journal} {J. Chem. Phys.}\ }\textbf {\bibinfo {volume} {145}},\ \bibinfo
  {pages} {104107} (\bibinfo {year} {2016})}\BibitemShut {NoStop}%
\bibitem [{\citenamefont {Konecny}\ \emph {et~al.}(2016)\citenamefont
  {Konecny}, \citenamefont {Kadek}, \citenamefont {Komorovsky}, \citenamefont
  {Malkina}, \citenamefont {Ruud},\ and\ \citenamefont
  {Repisky}}]{Repisky16_5823}%
  \BibitemOpen
  \bibfield  {author} {\bibinfo {author} {\bibfnamefont {L.}~\bibnamefont
  {Konecny}}, \bibinfo {author} {\bibfnamefont {M.}~\bibnamefont {Kadek}},
  \bibinfo {author} {\bibfnamefont {S.}~\bibnamefont {Komorovsky}}, \bibinfo
  {author} {\bibfnamefont {O.~L.}\ \bibnamefont {Malkina}}, \bibinfo {author}
  {\bibfnamefont {K.}~\bibnamefont {Ruud}},\ and\ \bibinfo {author}
  {\bibfnamefont {M.}~\bibnamefont {Repisky}},\ }\bibfield  {title} {\enquote
  {\bibinfo {title} {{Acceleration of Relativistic Electron Dynamics by Means
  of {X2C} Transformation: {A}pplication to the Calculation of Nonlinear
  Optical Properties}},}\ }\href {https://doi.org/10.1021/acs.jctc.6b00740}
  {\bibfield  {journal} {\bibinfo  {journal} {J. Chem. Theory Comput.}\
  }\textbf {\bibinfo {volume} {12}},\ \bibinfo {pages} {5823--5833} (\bibinfo
  {year} {2016})}\BibitemShut {NoStop}%
\bibitem [{\citenamefont {Egidi}\ \emph {et~al.}(2017)\citenamefont {Egidi},
  \citenamefont {Sun}, \citenamefont {Goings}, \citenamefont {Scalmani},
  \citenamefont {Frisch},\ and\ \citenamefont {Li}}]{Li17_2591}%
  \BibitemOpen
  \bibfield  {author} {\bibinfo {author} {\bibfnamefont {F.}~\bibnamefont
  {Egidi}}, \bibinfo {author} {\bibfnamefont {S.}~\bibnamefont {Sun}}, \bibinfo
  {author} {\bibfnamefont {J.~J.}\ \bibnamefont {Goings}}, \bibinfo {author}
  {\bibfnamefont {G.}~\bibnamefont {Scalmani}}, \bibinfo {author}
  {\bibfnamefont {M.~J.}\ \bibnamefont {Frisch}},\ and\ \bibinfo {author}
  {\bibfnamefont {X.}~\bibnamefont {Li}},\ }\bibfield  {title} {\enquote
  {\bibinfo {title} {Two-{{Component Noncollinear Time-Dependent Spin Density
  Functional Theory}} for {{Excited State Calculations}}},}\ }\href
  {https://doi.org/10.1021/acs.jctc.7b00104} {\bibfield  {journal} {\bibinfo
  {journal} {J. Chem. Theory Comput.}\ }\textbf {\bibinfo {volume} {13}},\
  \bibinfo {pages} {2591--2603} (\bibinfo {year} {2017})}\BibitemShut {NoStop}%
\bibitem [{\citenamefont {Shee}\ \emph {et~al.}(2018)\citenamefont {Shee},
  \citenamefont {Saue}, \citenamefont {Visscher},\ and\ \citenamefont {Severo
  Pereira~Gomes}}]{Gomes18_174113}%
  \BibitemOpen
  \bibfield  {author} {\bibinfo {author} {\bibfnamefont {A.}~\bibnamefont
  {Shee}}, \bibinfo {author} {\bibfnamefont {T.}~\bibnamefont {Saue}}, \bibinfo
  {author} {\bibfnamefont {L.}~\bibnamefont {Visscher}},\ and\ \bibinfo
  {author} {\bibfnamefont {A.}~\bibnamefont {Severo Pereira~Gomes}},\
  }\bibfield  {title} {\enquote {\bibinfo {title} {Equation-of-motion
  coupled-cluster theory based on the 4-component dirac-coulomb(-gaunt)
  hamiltonian. energies for single electron detachment, attachment, and
  electronically excited states},}\ }\href {https://doi.org/10.1063/1.5053846}
  {\bibfield  {journal} {\bibinfo  {journal} {J. Chem. Phys.}\ }\textbf
  {\bibinfo {volume} {149}},\ \bibinfo {pages} {174113} (\bibinfo {year}
  {2018})},\ \Eprint {https://arxiv.org/abs/https://doi.org/10.1063/1.5053846}
  {https://doi.org/10.1063/1.5053846} \BibitemShut {NoStop}%
\bibitem [{\citenamefont {Liu}\ \emph {et~al.}(2018)\citenamefont {Liu},
  \citenamefont {Shen}, \citenamefont {Asthana},\ and\ \citenamefont
  {Cheng}}]{Cheng18_034106}%
  \BibitemOpen
  \bibfield  {author} {\bibinfo {author} {\bibfnamefont {J.}~\bibnamefont
  {Liu}}, \bibinfo {author} {\bibfnamefont {Y.}~\bibnamefont {Shen}}, \bibinfo
  {author} {\bibfnamefont {A.}~\bibnamefont {Asthana}},\ and\ \bibinfo {author}
  {\bibfnamefont {L.}~\bibnamefont {Cheng}},\ }\bibfield  {title} {\enquote
  {\bibinfo {title} {{Two-component Relativistic Coupled-cluster Methods using
  Mean-field Spin-orbit Integrals}},}\ }\href
  {https://doi.org/10.1063/1.5009177} {\bibfield  {journal} {\bibinfo
  {journal} {J. Chem. Phys.}\ }\textbf {\bibinfo {volume} {148}},\ \bibinfo
  {pages} {034106} (\bibinfo {year} {2018})}\BibitemShut {NoStop}%
\bibitem [{\citenamefont {Asthana}, \citenamefont {Liu},\ and\ \citenamefont
  {Cheng}(2019)}]{Cheng19_074102}%
  \BibitemOpen
  \bibfield  {author} {\bibinfo {author} {\bibfnamefont {A.}~\bibnamefont
  {Asthana}}, \bibinfo {author} {\bibfnamefont {J.}~\bibnamefont {Liu}},\ and\
  \bibinfo {author} {\bibfnamefont {L.}~\bibnamefont {Cheng}},\ }\bibfield
  {title} {\enquote {\bibinfo {title} {Exact two-component equation-of-motion
  coupled-cluster singles and doubles method using atomic mean-field spin-orbit
  integrals},}\ }\href {https://doi.org/10.1063/1.5081715} {\bibfield
  {journal} {\bibinfo  {journal} {J. Chem. Phys.}\ }\textbf {\bibinfo {volume}
  {150}},\ \bibinfo {pages} {074102} (\bibinfo {year} {2019})}\BibitemShut
  {NoStop}%
\bibitem [{\citenamefont {Pototschnig}\ \emph {et~al.}(2021)\citenamefont
  {Pototschnig}, \citenamefont {Papadopoulos}, \citenamefont {Lyakh},
  \citenamefont {Repisky}, \citenamefont {Halbert}, \citenamefont {Severo
  Pereira~Gomes}, \citenamefont {Jensen},\ and\ \citenamefont
  {Visscher}}]{Visscher21_5509}%
  \BibitemOpen
  \bibfield  {author} {\bibinfo {author} {\bibfnamefont {J.~V.}\ \bibnamefont
  {Pototschnig}}, \bibinfo {author} {\bibfnamefont {A.}~\bibnamefont
  {Papadopoulos}}, \bibinfo {author} {\bibfnamefont {D.~I.}\ \bibnamefont
  {Lyakh}}, \bibinfo {author} {\bibfnamefont {M.}~\bibnamefont {Repisky}},
  \bibinfo {author} {\bibfnamefont {L.}~\bibnamefont {Halbert}}, \bibinfo
  {author} {\bibfnamefont {A.}~\bibnamefont {Severo Pereira~Gomes}}, \bibinfo
  {author} {\bibfnamefont {H.~J.~A.}\ \bibnamefont {Jensen}},\ and\ \bibinfo
  {author} {\bibfnamefont {L.}~\bibnamefont {Visscher}},\ }\bibfield  {title}
  {\enquote {\bibinfo {title} {Implementation of relativistic coupled cluster
  theory for massively parallel gpu-accelerated computing architectures},}\
  }\href {https://doi.org/10.1021/acs.jctc.1c00260} {\bibfield  {journal}
  {\bibinfo  {journal} {J. Chem. Theory Comput.}\ }\textbf {\bibinfo {volume}
  {17}},\ \bibinfo {pages} {5509--5529} (\bibinfo {year} {2021})}\BibitemShut
  {NoStop}%
\bibitem [{\citenamefont {Liu}\ and\ \citenamefont
  {Cheng}(2021)}]{Cheng21_e1536}%
  \BibitemOpen
  \bibfield  {author} {\bibinfo {author} {\bibfnamefont {J.}~\bibnamefont
  {Liu}}\ and\ \bibinfo {author} {\bibfnamefont {L.}~\bibnamefont {Cheng}},\
  }\bibfield  {title} {\enquote {\bibinfo {title} {Relativistic coupled-cluster
  and equation-of-motion coupled-cluster methods},}\ }\href
  {https://doi.org/10.1002/wcms.1536} {\bibfield  {journal} {\bibinfo
  {journal} {WIREs Comput. Mol. Sci.}\ }\textbf {\bibinfo {volume} {11}},\
  \bibinfo {pages} {e1536} (\bibinfo {year} {2021})}\BibitemShut {NoStop}%
\bibitem [{\citenamefont {Sharma}\ \emph {et~al.}(2022)\citenamefont {Sharma},
  \citenamefont {Jenkins}, \citenamefont {Scalmani}, \citenamefont {Frisch},
  \citenamefont {Truhlar}, \citenamefont {Gagliardi},\ and\ \citenamefont
  {Li}}]{Li22_2947}%
  \BibitemOpen
  \bibfield  {author} {\bibinfo {author} {\bibfnamefont {P.}~\bibnamefont
  {Sharma}}, \bibinfo {author} {\bibfnamefont {A.~J.}\ \bibnamefont {Jenkins}},
  \bibinfo {author} {\bibfnamefont {G.}~\bibnamefont {Scalmani}}, \bibinfo
  {author} {\bibfnamefont {M.~J.}\ \bibnamefont {Frisch}}, \bibinfo {author}
  {\bibfnamefont {D.~G.}\ \bibnamefont {Truhlar}}, \bibinfo {author}
  {\bibfnamefont {L.}~\bibnamefont {Gagliardi}},\ and\ \bibinfo {author}
  {\bibfnamefont {X.}~\bibnamefont {Li}},\ }\bibfield  {title} {\enquote
  {\bibinfo {title} {Exact-{{Two-Component Multiconfiguration Pair-Density
  Functional Theory}}},}\ }\href {https://doi.org/10.1021/acs.jctc.2c00062}
  {\bibfield  {journal} {\bibinfo  {journal} {J. Chem. Theory Comput.}\
  }\textbf {\bibinfo {volume} {18}},\ \bibinfo {pages} {2947--2954} (\bibinfo
  {year} {2022})}\BibitemShut {NoStop}%
\bibitem [{\citenamefont {Lu}\ \emph {et~al.}(2022)\citenamefont {Lu},
  \citenamefont {Hu}, \citenamefont {Jenkins},\ and\ \citenamefont
  {Li}}]{Li22_2983}%
  \BibitemOpen
  \bibfield  {author} {\bibinfo {author} {\bibfnamefont {L.}~\bibnamefont
  {Lu}}, \bibinfo {author} {\bibfnamefont {H.}~\bibnamefont {Hu}}, \bibinfo
  {author} {\bibfnamefont {A.~J.}\ \bibnamefont {Jenkins}},\ and\ \bibinfo
  {author} {\bibfnamefont {X.}~\bibnamefont {Li}},\ }\bibfield  {title}
  {\enquote {\bibinfo {title} {Exact-{{Two-Component Relativistic
  Multireference Second-Order Perturbation Theory}}},}\ }\href
  {https://doi.org/10.1021/acs.jctc.2c00171} {\bibfield  {journal} {\bibinfo
  {journal} {J. Chem. Theory Comput.}\ }\textbf {\bibinfo {volume} {18}},\
  \bibinfo {pages} {2983--2992} (\bibinfo {year} {2022})}\BibitemShut {NoStop}%
\bibitem [{\citenamefont {Hoyer}\ \emph {et~al.}(2022)\citenamefont {Hoyer},
  \citenamefont {Hu}, \citenamefont {Lu}, \citenamefont {Knecht},\ and\
  \citenamefont {Li}}]{Li22_5011}%
  \BibitemOpen
  \bibfield  {author} {\bibinfo {author} {\bibfnamefont {C.~E.}\ \bibnamefont
  {Hoyer}}, \bibinfo {author} {\bibfnamefont {H.}~\bibnamefont {Hu}}, \bibinfo
  {author} {\bibfnamefont {L.}~\bibnamefont {Lu}}, \bibinfo {author}
  {\bibfnamefont {S.}~\bibnamefont {Knecht}},\ and\ \bibinfo {author}
  {\bibfnamefont {X.}~\bibnamefont {Li}},\ }\bibfield  {title} {\enquote
  {\bibinfo {title} {Relativistic {{Kramers-Unrestricted Exact-Two-Component
  Density Matrix Renormalization Group}}},}\ }\href
  {https://doi.org/10.1021/acs.jpca.2c02150} {\bibfield  {journal} {\bibinfo
  {journal} {J. Phys. Chem. A}\ }\textbf {\bibinfo {volume} {126}},\ \bibinfo
  {pages} {5011--5020} (\bibinfo {year} {2022})}\BibitemShut {NoStop}%
\bibitem [{\citenamefont {Zhang}\ \emph {et~al.}(2024)\citenamefont {Zhang},
  \citenamefont {Banerjee}, \citenamefont {Koulias}, \citenamefont {Valeev},
  \citenamefont {DePrince~III},\ and\ \citenamefont {Li}}]{Li24_3408}%
  \BibitemOpen
  \bibfield  {author} {\bibinfo {author} {\bibfnamefont {T.}~\bibnamefont
  {Zhang}}, \bibinfo {author} {\bibfnamefont {S.}~\bibnamefont {Banerjee}},
  \bibinfo {author} {\bibfnamefont {L.~N.}\ \bibnamefont {Koulias}}, \bibinfo
  {author} {\bibfnamefont {E.~F.}\ \bibnamefont {Valeev}}, \bibinfo {author}
  {\bibfnamefont {A.~E.}\ \bibnamefont {DePrince~III}},\ and\ \bibinfo {author}
  {\bibfnamefont {X.}~\bibnamefont {Li}},\ }\bibfield  {title} {\enquote
  {\bibinfo {title} {Dirac--{{Coulomb}}--{{Breit Molecular Mean-Field
  Exact-Two-Component Relativistic Equation-of-Motion Coupled-Cluster
  Theory}}},}\ }\href {https://doi.org/10.1021/acs.jpca.3c08167} {\bibfield
  {journal} {\bibinfo  {journal} {J. Phys. Chem. A}\ }\textbf {\bibinfo
  {volume} {128}},\ \bibinfo {pages} {3408--3418} (\bibinfo {year}
  {2024})}\BibitemShut {NoStop}%
\bibitem [{\citenamefont {Kovtun}\ \emph {et~al.}(2024)\citenamefont {Kovtun},
  \citenamefont {Lambros}, \citenamefont {Liu}, \citenamefont {Tang},
  \citenamefont {Williams-Young},\ and\ \citenamefont {Li}}]{Li24_7694}%
  \BibitemOpen
  \bibfield  {author} {\bibinfo {author} {\bibfnamefont {M.}~\bibnamefont
  {Kovtun}}, \bibinfo {author} {\bibfnamefont {E.}~\bibnamefont {Lambros}},
  \bibinfo {author} {\bibfnamefont {A.}~\bibnamefont {Liu}}, \bibinfo {author}
  {\bibfnamefont {D.}~\bibnamefont {Tang}}, \bibinfo {author} {\bibfnamefont
  {D.~B.}\ \bibnamefont {Williams-Young}},\ and\ \bibinfo {author}
  {\bibfnamefont {X.}~\bibnamefont {Li}},\ }\bibfield  {title} {\enquote
  {\bibinfo {title} {Accelerating relativistic exact-two-component density
  functional theory calculations with graphical processing units},}\ }\href
  {https://doi.org/10.1021/acs.jctc.4c00843} {\bibfield  {journal} {\bibinfo
  {journal} {J. Chem. Theory Comput.}\ }\textbf {\bibinfo {volume} {20}},\
  \bibinfo {pages} {7694--7699} (\bibinfo {year} {2024})}\BibitemShut {NoStop}%
\bibitem [{\citenamefont {Hu}\ \emph {et~al.}(2024)\citenamefont {Hu},
  \citenamefont {Upadhyay}, \citenamefont {Lu}, \citenamefont {Jenkins},
  \citenamefont {Zhang}, \citenamefont {Shayit}, \citenamefont {Knecht},\ and\
  \citenamefont {Li}}]{Li24_041404}%
  \BibitemOpen
  \bibfield  {author} {\bibinfo {author} {\bibfnamefont {H.}~\bibnamefont
  {Hu}}, \bibinfo {author} {\bibfnamefont {S.}~\bibnamefont {Upadhyay}},
  \bibinfo {author} {\bibfnamefont {L.}~\bibnamefont {Lu}}, \bibinfo {author}
  {\bibfnamefont {A.~J.}\ \bibnamefont {Jenkins}}, \bibinfo {author}
  {\bibfnamefont {T.}~\bibnamefont {Zhang}}, \bibinfo {author} {\bibfnamefont
  {A.}~\bibnamefont {Shayit}}, \bibinfo {author} {\bibfnamefont
  {S.}~\bibnamefont {Knecht}},\ and\ \bibinfo {author} {\bibfnamefont
  {X.}~\bibnamefont {Li}},\ }\bibfield  {title} {\enquote {\bibinfo {title}
  {{Small Tensor Product Distributed Active Space (STP-DAS) Framework for
  Relativistic and Non-relativistic Multiconfiguration Calculations: Scaling
  from 10$^9$ on a Laptop to 10$^{12}$ Determinants on a Supercomputer}},}\
  }\href {https://doi.org/10.1063/5.0227122} {\bibfield  {journal} {\bibinfo
  {journal} {Comput. Phys. Rep.}\ }\textbf {\bibinfo {volume} {5}},\ \bibinfo
  {pages} {041404} (\bibinfo {year} {2024})}\BibitemShut {NoStop}%
\bibitem [{\citenamefont {Boettger}(2000)}]{Boettger00_7809}%
  \BibitemOpen
  \bibfield  {author} {\bibinfo {author} {\bibfnamefont {J.~C.}\ \bibnamefont
  {Boettger}},\ }\bibfield  {title} {\enquote {\bibinfo {title} {Approximate
  two-electron spin-orbit coupling term for density-functional-theory {DFT}
  calculations using the {D}ouglas-{K}roll-{H}ess transformation},}\ }\href
  {https://doi.org/10.1103/PhysRevB.62.7809} {\bibfield  {journal} {\bibinfo
  {journal} {Phys. Rev. B}\ }\textbf {\bibinfo {volume} {62}},\ \bibinfo
  {pages} {7809--7815} (\bibinfo {year} {2000})}\BibitemShut {NoStop}%
\bibitem [{\citenamefont {Sun}\ \emph {et~al.}(2021)\citenamefont {Sun},
  \citenamefont {Stetina}, \citenamefont {Zhang}, \citenamefont {Hu},
  \citenamefont {Valeev}, \citenamefont {Sun},\ and\ \citenamefont
  {Li}}]{Li21_3388}%
  \BibitemOpen
  \bibfield  {author} {\bibinfo {author} {\bibfnamefont {S.}~\bibnamefont
  {Sun}}, \bibinfo {author} {\bibfnamefont {T.~F.}\ \bibnamefont {Stetina}},
  \bibinfo {author} {\bibfnamefont {T.}~\bibnamefont {Zhang}}, \bibinfo
  {author} {\bibfnamefont {H.}~\bibnamefont {Hu}}, \bibinfo {author}
  {\bibfnamefont {E.~F.}\ \bibnamefont {Valeev}}, \bibinfo {author}
  {\bibfnamefont {Q.}~\bibnamefont {Sun}},\ and\ \bibinfo {author}
  {\bibfnamefont {X.}~\bibnamefont {Li}},\ }\bibfield  {title} {\enquote
  {\bibinfo {title} {Efficient four-component dirac--coulomb--gaunt
  hartree--fock in the pauli spinor representation},}\ }\href
  {https://doi.org/10.1021/acs.jctc.1c00137} {\bibfield  {journal} {\bibinfo
  {journal} {J. Chem. Theory Comput.}\ }\textbf {\bibinfo {volume} {17}},\
  \bibinfo {pages} {3388--3402} (\bibinfo {year} {2021})}\BibitemShut {NoStop}%
\bibitem [{\citenamefont {Sun}\ \emph {et~al.}(2022)\citenamefont {Sun},
  \citenamefont {Ehrman}, \citenamefont {Sun},\ and\ \citenamefont
  {Li}}]{Li22_064112}%
  \BibitemOpen
  \bibfield  {author} {\bibinfo {author} {\bibfnamefont {S.}~\bibnamefont
  {Sun}}, \bibinfo {author} {\bibfnamefont {J.~N.}\ \bibnamefont {Ehrman}},
  \bibinfo {author} {\bibfnamefont {Q.}~\bibnamefont {Sun}},\ and\ \bibinfo
  {author} {\bibfnamefont {X.}~\bibnamefont {Li}},\ }\bibfield  {title}
  {\enquote {\bibinfo {title} {Efficient evaluation of the breit operator in
  the pauli spinor basis},}\ }\href {https://doi.org/10.1063/5.0098828}
  {\bibfield  {journal} {\bibinfo  {journal} {J. Chem. Phys.}\ }\textbf
  {\bibinfo {volume} {157}},\ \bibinfo {pages} {064112} (\bibinfo {year}
  {2022})}\BibitemShut {NoStop}%
\bibitem [{\citenamefont {Wang}\ \emph {et~al.}(2015)\citenamefont {Wang},
  \citenamefont {Hu}, \citenamefont {Wang},\ and\ \citenamefont
  {Guo}}]{Guo15_144109}%
  \BibitemOpen
  \bibfield  {author} {\bibinfo {author} {\bibfnamefont {Z.}~\bibnamefont
  {Wang}}, \bibinfo {author} {\bibfnamefont {S.}~\bibnamefont {Hu}}, \bibinfo
  {author} {\bibfnamefont {F.}~\bibnamefont {Wang}},\ and\ \bibinfo {author}
  {\bibfnamefont {J.}~\bibnamefont {Guo}},\ }\bibfield  {title} {\enquote
  {\bibinfo {title} {{Equation-of-motion coupled-cluster method for doubly
  ionized states with spin-orbit coupling}},}\ }\href
  {https://doi.org/10.1063/1.4917041} {\bibfield  {journal} {\bibinfo
  {journal} {J. Chem. Phys.}\ }\textbf {\bibinfo {volume} {142}},\ \bibinfo
  {pages} {144109} (\bibinfo {year} {2015})}\BibitemShut {NoStop}%
\bibitem [{\citenamefont {Zhao}\ \emph {et~al.}(2020)\citenamefont {Zhao},
  \citenamefont {Wang}, \citenamefont {Guo},\ and\ \citenamefont
  {Wang}}]{wang20_134105}%
  \BibitemOpen
  \bibfield  {author} {\bibinfo {author} {\bibfnamefont {H.}~\bibnamefont
  {Zhao}}, \bibinfo {author} {\bibfnamefont {Z.}~\bibnamefont {Wang}}, \bibinfo
  {author} {\bibfnamefont {M.}~\bibnamefont {Guo}},\ and\ \bibinfo {author}
  {\bibfnamefont {F.}~\bibnamefont {Wang}},\ }\bibfield  {title} {\enquote
  {\bibinfo {title} {{Splittings of d8 configurations of late-transition metals
  with EOM-DIP-CCSD and FSCCSD methods}},}\ }\href
  {https://doi.org/10.1063/1.5145077} {\bibfield  {journal} {\bibinfo
  {journal} {J. Chem. Phys.}\ }\textbf {\bibinfo {volume} {152}},\ \bibinfo
  {pages} {134105} (\bibinfo {year} {2020})}\BibitemShut {NoStop}%
\bibitem [{\citenamefont {Pathak}\ \emph {et~al.}(2014)\citenamefont {Pathak},
  \citenamefont {Sasmal}, \citenamefont {Nayak}, \citenamefont {Vaval},\ and\
  \citenamefont {Pal}}]{Pal14_062501}%
  \BibitemOpen
  \bibfield  {author} {\bibinfo {author} {\bibfnamefont {H.}~\bibnamefont
  {Pathak}}, \bibinfo {author} {\bibfnamefont {S.}~\bibnamefont {Sasmal}},
  \bibinfo {author} {\bibfnamefont {M.~K.}\ \bibnamefont {Nayak}}, \bibinfo
  {author} {\bibfnamefont {N.}~\bibnamefont {Vaval}},\ and\ \bibinfo {author}
  {\bibfnamefont {S.}~\bibnamefont {Pal}},\ }\bibfield  {title} {\enquote
  {\bibinfo {title} {Relativistic equation-of-motion coupled-cluster method for
  the ionization problem: Application to molecules},}\ }\href
  {https://doi.org/10.1103/PhysRevA.90.062501} {\bibfield  {journal} {\bibinfo
  {journal} {Phys. Rev. A}\ }\textbf {\bibinfo {volume} {90}},\ \bibinfo
  {pages} {062501} (\bibinfo {year} {2014})}\BibitemShut {NoStop}%
\bibitem [{\citenamefont {Pathak}\ \emph {et~al.}(2020)\citenamefont {Pathak},
  \citenamefont {Sasmal}, \citenamefont {Talukdar}, \citenamefont {Nayak},
  \citenamefont {Vaval},\ and\ \citenamefont {Pal}}]{Pal20_104302}%
  \BibitemOpen
  \bibfield  {author} {\bibinfo {author} {\bibfnamefont {H.}~\bibnamefont
  {Pathak}}, \bibinfo {author} {\bibfnamefont {S.}~\bibnamefont {Sasmal}},
  \bibinfo {author} {\bibfnamefont {K.}~\bibnamefont {Talukdar}}, \bibinfo
  {author} {\bibfnamefont {M.~K.}\ \bibnamefont {Nayak}}, \bibinfo {author}
  {\bibfnamefont {N.}~\bibnamefont {Vaval}},\ and\ \bibinfo {author}
  {\bibfnamefont {S.}~\bibnamefont {Pal}},\ }\bibfield  {title} {\enquote
  {\bibinfo {title} {Relativistic double-ionization equation-of-motion
  coupled-cluster method: {{Application}} to low-lying doubly ionized
  states},}\ }\href {https://doi.org/10.1063/1.5140988} {\bibfield  {journal}
  {\bibinfo  {journal} {J. Chem. Phys.}\ }\textbf {\bibinfo {volume} {152}},\
  \bibinfo {pages} {104302} (\bibinfo {year} {2020})}\BibitemShut {NoStop}%
\bibitem [{\citenamefont {Ehrman}\ \emph {et~al.}(2023)\citenamefont {Ehrman},
  \citenamefont {{Martinez-Baez}}, \citenamefont {Jenkins},\ and\ \citenamefont
  {Li}}]{Li23_5785}%
  \BibitemOpen
  \bibfield  {author} {\bibinfo {author} {\bibfnamefont {J.}~\bibnamefont
  {Ehrman}}, \bibinfo {author} {\bibfnamefont {E.}~\bibnamefont
  {{Martinez-Baez}}}, \bibinfo {author} {\bibfnamefont {A.~J.}\ \bibnamefont
  {Jenkins}},\ and\ \bibinfo {author} {\bibfnamefont {X.}~\bibnamefont {Li}},\
  }\bibfield  {title} {\enquote {\bibinfo {title} {Improving {{One-Electron
  Exact-Two-Component Relativistic Methods}} with the
  {{Dirac}}--{{Coulomb}}--{{Breit-Parameterized Effective Spin}}--{{Orbit
  Coupling}}},}\ }\href {https://doi.org/10.1021/acs.jctc.3c00479} {\bibfield
  {journal} {\bibinfo  {journal} {J. Chem. Theory Comput.}\ }\textbf {\bibinfo
  {volume} {19}},\ \bibinfo {pages} {5785--5790} (\bibinfo {year}
  {2023})}\BibitemShut {NoStop}%
\bibitem [{\citenamefont {Yuwono}\ \emph {et~al.}(2025)\citenamefont {Yuwono},
  \citenamefont {Li}, \citenamefont {Zhang}, \citenamefont {Li},\ and\
  \citenamefont {DePrince}}]{DePrince25_084110}%
  \BibitemOpen
  \bibfield  {author} {\bibinfo {author} {\bibfnamefont {S.~H.}\ \bibnamefont
  {Yuwono}}, \bibinfo {author} {\bibfnamefont {R.~R.}\ \bibnamefont {Li}},
  \bibinfo {author} {\bibfnamefont {T.}~\bibnamefont {Zhang}}, \bibinfo
  {author} {\bibfnamefont {X.}~\bibnamefont {Li}},\ and\ \bibinfo {author}
  {\bibfnamefont {I.}~\bibnamefont {DePrince}, \bibfnamefont {A.~Eugene}},\
  }\bibfield  {title} {\enquote {\bibinfo {title} {Two-component relativistic
  equation-of-motion coupled cluster for electron ionization},}\ }\href
  {https://doi.org/10.1063/5.0248535} {\bibfield  {journal} {\bibinfo
  {journal} {J. Chem. Phys.}\ }\textbf {\bibinfo {volume} {162}},\ \bibinfo
  {pages} {084110} (\bibinfo {year} {2025})}\BibitemShut {NoStop}%
\bibitem [{\citenamefont {Shen}\ and\ \citenamefont
  {Piecuch}(2021)}]{Piecuch21_e1966534}%
  \BibitemOpen
  \bibfield  {author} {\bibinfo {author} {\bibfnamefont {J.}~\bibnamefont
  {Shen}}\ and\ \bibinfo {author} {\bibfnamefont {P.}~\bibnamefont {Piecuch}},\
  }\bibfield  {title} {\enquote {\bibinfo {title} {Double electron-attachment
  equation-of-motion coupled-cluster methods with up to 4-particle--2-hole
  excitations: Improved implementation and application to singlet--triplet gaps
  in ortho-, meta-, and para-benzyne isomers},}\ }\href
  {https://doi.org/10.1080/00268976.2021.1966534} {\bibfield  {journal}
  {\bibinfo  {journal} {Mol. Phys.}\ }\textbf {\bibinfo {volume} {119}},\
  \bibinfo {pages} {e1966534} (\bibinfo {year} {2021})}\BibitemShut {NoStop}%
\bibitem [{\citenamefont {Pollak}\ and\ \citenamefont
  {Weigend}(2017)}]{Weigend17_3696}%
  \BibitemOpen
  \bibfield  {author} {\bibinfo {author} {\bibfnamefont {P.}~\bibnamefont
  {Pollak}}\ and\ \bibinfo {author} {\bibfnamefont {F.}~\bibnamefont
  {Weigend}},\ }\bibfield  {title} {\enquote {\bibinfo {title} {Segmented
  {{Contracted Error-Consistent Basis Sets}} of {{Double-}} and
  {{Triple-$\zeta$ Valence Quality}} for {{One-}} and {{Two-Component
  Relativistic All-Electron Calculations}}},}\ }\href
  {https://doi.org/10.1021/acs.jctc.7b00593} {\bibfield  {journal} {\bibinfo
  {journal} {J. Chem. Theory Comput.}\ }\textbf {\bibinfo {volume} {13}},\
  \bibinfo {pages} {3696--3705} (\bibinfo {year} {2017})}\BibitemShut {NoStop}%
\bibitem [{\citenamefont {Dyall}(2002)}]{Dyall02_335}%
  \BibitemOpen
  \bibfield  {author} {\bibinfo {author} {\bibfnamefont {K.~G.}\ \bibnamefont
  {Dyall}},\ }\bibfield  {title} {\enquote {\bibinfo {title} {Relativistic and
  nonrelativistic finite nucleus optimized triple-zeta basis sets for the 4p,
  5p and 6p elements},}\ }\href {https://doi.org/10.1007/s00214-002-0388-0}
  {\bibfield  {journal} {\bibinfo  {journal} {Theor. Chem. Acc.}\ }\textbf
  {\bibinfo {volume} {108}},\ \bibinfo {pages} {335--340} (\bibinfo {year}
  {2002})},\ \bibinfo {note}
  {\href{https://doi.org/10.1007/s00214-003-0433-7}{{\bf 109}, 284 (2003)
  [Erratum]}}\BibitemShut {NoStop}%
\bibitem [{\citenamefont {Dyall}(2006)}]{Dyall06_441}%
  \BibitemOpen
  \bibfield  {author} {\bibinfo {author} {\bibfnamefont {K.~G.}\ \bibnamefont
  {Dyall}},\ }\bibfield  {title} {\enquote {\bibinfo {title} {Relativistic
  {{Quadruple-Zeta}} and {{Revised Triple-Zeta}} and {{Double-Zeta Basis Sets}}
  for the 4p, 5p, and 6p {{Elements}}},}\ }\href
  {https://doi.org/10.1007/s00214-006-0126-0} {\bibfield  {journal} {\bibinfo
  {journal} {Theor. Chem. Acc.}\ }\textbf {\bibinfo {volume} {115}},\ \bibinfo
  {pages} {441--447} (\bibinfo {year} {2006})}\BibitemShut {NoStop}%
\bibitem [{\citenamefont {Dyall}(2016)}]{Dyall16_128}%
  \BibitemOpen
  \bibfield  {author} {\bibinfo {author} {\bibfnamefont {K.~G.}\ \bibnamefont
  {Dyall}},\ }\bibfield  {title} {\enquote {\bibinfo {title} {Relativistic
  double-zeta, triple-zeta, and quadruple-zeta basis sets for the light
  elements {{H}}--{{Ar}}},}\ }\href {https://doi.org/10.1007/s00214-016-1884-y}
  {\bibfield  {journal} {\bibinfo  {journal} {Theor. Chem. Acc.}\ }\textbf
  {\bibinfo {volume} {135}},\ \bibinfo {pages} {128} (\bibinfo {year}
  {2016})}\BibitemShut {NoStop}%
\bibitem [{\citenamefont {Dyall}(2012{\natexlab{a}})}]{Dyall12_1217}%
  \BibitemOpen
  \bibfield  {author} {\bibinfo {author} {\bibfnamefont {K.~G.}\ \bibnamefont
  {Dyall}},\ }\bibfield  {title} {\enquote {\bibinfo {title} {Core correlating
  basis functions for elements 31--118},}\ }\href
  {https://doi.org/10.1007/s00214-012-1217-8} {\bibfield  {journal} {\bibinfo
  {journal} {Theor Chem Acc}\ }\textbf {\bibinfo {volume} {131}},\ \bibinfo
  {pages} {1217} (\bibinfo {year} {2012}{\natexlab{a}})}\BibitemShut {NoStop}%
\bibitem [{\citenamefont {Dyall}(2012{\natexlab{b}})}]{Dyall12_1172}%
  \BibitemOpen
  \bibfield  {author} {\bibinfo {author} {\bibfnamefont {K.~G.}\ \bibnamefont
  {Dyall}},\ }\bibfield  {title} {\enquote {\bibinfo {title} {Relativistic
  double-zeta, triple-zeta, and quadruple-zeta basis sets for the 7p elements,
  with atomic and molecular applications},}\ }\href@noop {} {\bibfield
  {journal} {\bibinfo  {journal} {Theor. Chem. Acc.}\ }\textbf {\bibinfo
  {volume} {131}},\ \bibinfo {pages} {1--20} (\bibinfo {year}
  {2012}{\natexlab{b}})}\BibitemShut {NoStop}%
\bibitem [{\citenamefont {Dyall}()}]{DyallBasisZenodo}%
  \BibitemOpen
  \bibfield  {author} {\bibinfo {author} {\bibfnamefont {K.~G.}\ \bibnamefont
  {Dyall}},\ }\href {https://doi.org/10.5281/ZENODO.7574628} {\enquote
  {\bibinfo {title} {Dyall dz, tz, and qz basis sets for relativistic
  electronic structure calculations},}\ }\bibinfo {note} {Last accessed
  November 11, 2024}\BibitemShut {NoStop}%
\bibitem [{\citenamefont {Roos}\ \emph {et~al.}(2005)\citenamefont {Roos},
  \citenamefont {Lindh}, \citenamefont {Malmqvist}, \citenamefont {Veryazov},\
  and\ \citenamefont {Widmark}}]{Widmark05_6575}%
  \BibitemOpen
  \bibfield  {author} {\bibinfo {author} {\bibfnamefont {B.~O.}\ \bibnamefont
  {Roos}}, \bibinfo {author} {\bibfnamefont {R.}~\bibnamefont {Lindh}},
  \bibinfo {author} {\bibfnamefont {P.-{\AA}.}\ \bibnamefont {Malmqvist}},
  \bibinfo {author} {\bibfnamefont {V.}~\bibnamefont {Veryazov}},\ and\
  \bibinfo {author} {\bibfnamefont {P.-O.}\ \bibnamefont {Widmark}},\
  }\bibfield  {title} {\enquote {\bibinfo {title} {New {{Relativistic ANO Basis
  Sets}} for {{Transition Metal Atoms}}},}\ }\href
  {https://doi.org/10.1021/jp0581126} {\bibfield  {journal} {\bibinfo
  {journal} {J. Phys. Chem. A}\ }\textbf {\bibinfo {volume} {109}},\ \bibinfo
  {pages} {6575--6579} (\bibinfo {year} {2005})}\BibitemShut {NoStop}%
\bibitem [{\citenamefont {Roos}\ \emph {et~al.}(2004)\citenamefont {Roos},
  \citenamefont {Lindh}, \citenamefont {Malmqvist}, \citenamefont {Veryazov},\
  and\ \citenamefont {Widmark}}]{roos04_2851}%
  \BibitemOpen
  \bibfield  {author} {\bibinfo {author} {\bibfnamefont {B.~O.}\ \bibnamefont
  {Roos}}, \bibinfo {author} {\bibfnamefont {R.}~\bibnamefont {Lindh}},
  \bibinfo {author} {\bibfnamefont {P.-{\AA}.}\ \bibnamefont {Malmqvist}},
  \bibinfo {author} {\bibfnamefont {V.}~\bibnamefont {Veryazov}},\ and\
  \bibinfo {author} {\bibfnamefont {P.-O.}\ \bibnamefont {Widmark}},\
  }\bibfield  {title} {\enquote {\bibinfo {title} {Main group atoms and dimers
  studied with a new relativistic ano basis set},}\ }\href
  {https://doi.org/10.1021/jp031064+} {\bibfield  {journal} {\bibinfo
  {journal} {J. Phys. Chem. A}\ }\textbf {\bibinfo {volume} {108}},\ \bibinfo
  {pages} {2851--2858} (\bibinfo {year} {2004})}\BibitemShut {NoStop}%
\bibitem [{\citenamefont {{DePrince III}}\ and\ \citenamefont
  {Sherrill}(2013)}]{Sherrill13_2687}%
  \BibitemOpen
  \bibfield  {author} {\bibinfo {author} {\bibfnamefont {A.~E.}\ \bibnamefont
  {{DePrince III}}}\ and\ \bibinfo {author} {\bibfnamefont {C.~D.}\
  \bibnamefont {Sherrill}},\ }\bibfield  {title} {\enquote {\bibinfo {title}
  {Accuracy and efficiency of coupled-cluster theory using density
  fitting/cholesky decomposition, frozen natural orbitals, and a t1-transformed
  hamiltonian},}\ }\href {https://doi.org/10.1021/ct400250u} {\bibfield
  {journal} {\bibinfo  {journal} {J. Chem. Theory Comput.}\ }\textbf {\bibinfo
  {volume} {9}},\ \bibinfo {pages} {2687--2696} (\bibinfo {year}
  {2013})}\BibitemShut {NoStop}%
\bibitem [{\citenamefont {Williams-Young}\ \emph {et~al.}(2020)\citenamefont
  {Williams-Young}, \citenamefont {Petrone}, \citenamefont {Sun}, \citenamefont
  {Stetina}, \citenamefont {Lestrange}, \citenamefont {Hoyer}, \citenamefont
  {Nascimento}, \citenamefont {Koulias}, \citenamefont {Wildman}, \citenamefont
  {Kasper}, \citenamefont {Goings}, \citenamefont {Ding}, \citenamefont
  {DePrince~III}, \citenamefont {Valeev},\ and\ \citenamefont
  {Li}}]{Li20_e1436}%
  \BibitemOpen
  \bibfield  {author} {\bibinfo {author} {\bibfnamefont {D.~B.}\ \bibnamefont
  {Williams-Young}}, \bibinfo {author} {\bibfnamefont {A.}~\bibnamefont
  {Petrone}}, \bibinfo {author} {\bibfnamefont {S.}~\bibnamefont {Sun}},
  \bibinfo {author} {\bibfnamefont {T.~F.}\ \bibnamefont {Stetina}}, \bibinfo
  {author} {\bibfnamefont {P.}~\bibnamefont {Lestrange}}, \bibinfo {author}
  {\bibfnamefont {C.~E.}\ \bibnamefont {Hoyer}}, \bibinfo {author}
  {\bibfnamefont {D.~R.}\ \bibnamefont {Nascimento}}, \bibinfo {author}
  {\bibfnamefont {L.}~\bibnamefont {Koulias}}, \bibinfo {author} {\bibfnamefont
  {A.}~\bibnamefont {Wildman}}, \bibinfo {author} {\bibfnamefont
  {J.}~\bibnamefont {Kasper}}, \bibinfo {author} {\bibfnamefont {J.~J.}\
  \bibnamefont {Goings}}, \bibinfo {author} {\bibfnamefont {F.}~\bibnamefont
  {Ding}}, \bibinfo {author} {\bibfnamefont {A.~E.}\ \bibnamefont
  {DePrince~III}}, \bibinfo {author} {\bibfnamefont {E.~F.}\ \bibnamefont
  {Valeev}},\ and\ \bibinfo {author} {\bibfnamefont {X.}~\bibnamefont {Li}},\
  }\bibfield  {title} {\enquote {\bibinfo {title} {The chronus quantum software
  package},}\ }\href {https://doi.org/10.1002/wcms.1436} {\bibfield  {journal}
  {\bibinfo  {journal} {WIRES Comput. Mol. Sci.}\ }\textbf {\bibinfo {volume}
  {10}},\ \bibinfo {pages} {e1436} (\bibinfo {year} {2020})}\BibitemShut
  {NoStop}%
\bibitem [{\citenamefont {Calvin}\ and\ \citenamefont {Valeev}()}]{TiledArray}%
  \BibitemOpen
  \bibfield  {author} {\bibinfo {author} {\bibfnamefont {J.~A.}\ \bibnamefont
  {Calvin}}\ and\ \bibinfo {author} {\bibfnamefont {E.~F.}\ \bibnamefont
  {Valeev}},\ }\href {https://github.com/valeevgroup/tiledarray} {\enquote
  {\bibinfo {title} {Tiledarray: A general-purpose scalable block-sparse tensor
  framework},}\ }\bibinfo {note} {Last accessed June 10, 2024.}\BibitemShut
  {Stop}%
\bibitem [{\citenamefont {Rubin}\ and\ \citenamefont {{DePrince
  III}}(2021)}]{DePrince21_e1954709}%
  \BibitemOpen
  \bibfield  {author} {\bibinfo {author} {\bibfnamefont {N.~C.}\ \bibnamefont
  {Rubin}}\ and\ \bibinfo {author} {\bibfnamefont {A.~E.}\ \bibnamefont
  {{DePrince III}}},\ }\bibfield  {title} {\enquote {\bibinfo {title} {p†q: a
  tool for prototyping many-body methods for quantum chemistry},}\ }\href
  {https://doi.org/10.1080/00268976.2021.1954709} {\bibfield  {journal}
  {\bibinfo  {journal} {Mol. Phys.}\ }\textbf {\bibinfo {volume} {119}},\
  \bibinfo {pages} {e1954709} (\bibinfo {year} {2021})}\BibitemShut {NoStop}%
\bibitem [{\citenamefont {Liebenthal}\ \emph {et~al.}(2025)\citenamefont
  {Liebenthal}, \citenamefont {Yuwono}, \citenamefont {Koulias}, \citenamefont
  {Li}, \citenamefont {Rubin},\ and\ \citenamefont {{DePrince
  III}}}]{DePrince25_2501.08882}%
  \BibitemOpen
  \bibfield  {author} {\bibinfo {author} {\bibfnamefont {M.~D.}\ \bibnamefont
  {Liebenthal}}, \bibinfo {author} {\bibfnamefont {S.~H.}\ \bibnamefont
  {Yuwono}}, \bibinfo {author} {\bibfnamefont {L.~N.}\ \bibnamefont {Koulias}},
  \bibinfo {author} {\bibfnamefont {R.~R.}\ \bibnamefont {Li}}, \bibinfo
  {author} {\bibfnamefont {N.~C.}\ \bibnamefont {Rubin}},\ and\ \bibinfo
  {author} {\bibfnamefont {A.~E.}\ \bibnamefont {{DePrince III}}},\ }\bibfield
  {title} {\enquote {\bibinfo {title} {Automated quantum chemistry code
  generation with the p$^\dagger$q package},}\ }\href
  {https://arxiv.org/abs/2501.08882} {\bibfield  {journal} {\bibinfo  {journal}
  {arXiv preprint}\ ,\ \bibinfo {pages} {2501.08882}} (\bibinfo {year}
  {2025})}\BibitemShut {NoStop}%
\bibitem [{\citenamefont {Gururangan}, \citenamefont {Deustua},\ and\
  \citenamefont {Piecuch}()}]{CCpy}%
  \BibitemOpen
  \bibfield  {author} {\bibinfo {author} {\bibfnamefont {K.}~\bibnamefont
  {Gururangan}}, \bibinfo {author} {\bibfnamefont {J.~E.}\ \bibnamefont
  {Deustua}},\ and\ \bibinfo {author} {\bibfnamefont {P.}~\bibnamefont
  {Piecuch}},\ }\href {https://github.com/piecuch-group/ccpy} {\enquote
  {\bibinfo {title} {{CC}py: A coupled-cluster package written in {P}ython},}\
  }\bibinfo {note} {Last accessed November 11, 2024}\BibitemShut {NoStop}%
\bibitem [{\citenamefont {Kramida}\ \emph {et~al.}(2024)\citenamefont
  {Kramida}, \citenamefont {{Yu.~Ralchenko}}, \citenamefont {Reader},\ and\
  \citenamefont {{and NIST ASD Team}}}]{NIST}%
  \BibitemOpen
  \bibfield  {author} {\bibinfo {author} {\bibfnamefont {A.}~\bibnamefont
  {Kramida}}, \bibinfo {author} {\bibnamefont {{Yu.~Ralchenko}}}, \bibinfo
  {author} {\bibfnamefont {J.}~\bibnamefont {Reader}},\ and\ \bibinfo {author}
  {\bibnamefont {{and NIST ASD Team}}},\ }\href@noop {} {}\bibinfo
  {howpublished} {{NIST Atomic Spectra Database (ver. 5.12), [Online].
  Available: {\tt{https://physics.nist.gov/asd}} [2025, March 18]. National
  Institute of Standards and Technology, Gaithersburg, MD.}} (\bibinfo {year}
  {2024})\BibitemShut {NoStop}%
\bibitem [{\citenamefont {McConkey}\ \emph {et~al.}(1994)\citenamefont
  {McConkey}, \citenamefont {Dawber}, \citenamefont {Avaldi}, \citenamefont
  {MacDonald}, \citenamefont {King},\ and\ \citenamefont {Hall}}]{Hall94_271}%
  \BibitemOpen
  \bibfield  {author} {\bibinfo {author} {\bibfnamefont {A.~G.}\ \bibnamefont
  {McConkey}}, \bibinfo {author} {\bibfnamefont {G.}~\bibnamefont {Dawber}},
  \bibinfo {author} {\bibfnamefont {L.}~\bibnamefont {Avaldi}}, \bibinfo
  {author} {\bibfnamefont {M.~A.}\ \bibnamefont {MacDonald}}, \bibinfo {author}
  {\bibfnamefont {G.~C.}\ \bibnamefont {King}},\ and\ \bibinfo {author}
  {\bibfnamefont {R.~I.}\ \bibnamefont {Hall}},\ }\bibfield  {title} {\enquote
  {\bibinfo {title} {Threshold photoelectrons coincidence spectroscopy of
  doubly charged ions of hydrogen chloride and chlorine},}\ }\href
  {https://doi.org/10.1088/0953-4075/27/2/005} {\bibfield  {journal} {\bibinfo
  {journal} {J. Phys. B: At. Mol. Opt. Phys.}\ }\textbf {\bibinfo {volume}
  {27}},\ \bibinfo {pages} {271} (\bibinfo {year} {1994})}\BibitemShut
  {NoStop}%
\bibitem [{\citenamefont {Fleig}\ \emph {et~al.}(2008)\citenamefont {Fleig},
  \citenamefont {Edvardsson}, \citenamefont {Banks},\ and\ \citenamefont
  {Eland}}]{Eland08_270}%
  \BibitemOpen
  \bibfield  {author} {\bibinfo {author} {\bibfnamefont {T.}~\bibnamefont
  {Fleig}}, \bibinfo {author} {\bibfnamefont {D.}~\bibnamefont {Edvardsson}},
  \bibinfo {author} {\bibfnamefont {S.~T.}\ \bibnamefont {Banks}},\ and\
  \bibinfo {author} {\bibfnamefont {J.~H.}\ \bibnamefont {Eland}},\ }\bibfield
  {title} {\enquote {\bibinfo {title} {A theoretical and experimental study of
  the double photoionisation of molecular bromine and a new double ionisation
  mechanism},}\ }\href
  {https://doi.org/https://doi.org/10.1016/j.chemphys.2007.08.007} {\bibfield
  {journal} {\bibinfo  {journal} {Chem. Phys.}\ }\textbf {\bibinfo {volume}
  {343}},\ \bibinfo {pages} {270--280} (\bibinfo {year} {2008})}\BibitemShut
  {NoStop}%
\bibitem [{\citenamefont {Eland}(2003)}]{Eland03_171}%
  \BibitemOpen
  \bibfield  {author} {\bibinfo {author} {\bibfnamefont {J.~H.}\ \bibnamefont
  {Eland}},\ }\bibfield  {title} {\enquote {\bibinfo {title} {Complete double
  photoionisation spectra of small molecules from tof-pepeco measurements},}\
  }\href {https://doi.org/https://doi.org/10.1016/j.chemphys.2003.08.001}
  {\bibfield  {journal} {\bibinfo  {journal} {Chem. Phys.}\ }\textbf {\bibinfo
  {volume} {294}},\ \bibinfo {pages} {171--186} (\bibinfo {year}
  {2003})}\BibitemShut {NoStop}%
\bibitem [{\citenamefont {Yencha}\ \emph {et~al.}(2004)\citenamefont {Yencha},
  \citenamefont {Juarez}, \citenamefont {{Pui Lee}}, \citenamefont {King},
  \citenamefont {Bennett}, \citenamefont {Kemp},\ and\ \citenamefont
  {McNab}}]{McNab04_179}%
  \BibitemOpen
  \bibfield  {author} {\bibinfo {author} {\bibfnamefont {A.~J.}\ \bibnamefont
  {Yencha}}, \bibinfo {author} {\bibfnamefont {A.~M.}\ \bibnamefont {Juarez}},
  \bibinfo {author} {\bibfnamefont {S.}~\bibnamefont {{Pui Lee}}}, \bibinfo
  {author} {\bibfnamefont {G.~C.}\ \bibnamefont {King}}, \bibinfo {author}
  {\bibfnamefont {F.~R.}\ \bibnamefont {Bennett}}, \bibinfo {author}
  {\bibfnamefont {F.}~\bibnamefont {Kemp}},\ and\ \bibinfo {author}
  {\bibfnamefont {I.~R.}\ \bibnamefont {McNab}},\ }\bibfield  {title} {\enquote
  {\bibinfo {title} {Photo-double ionization of hydrogen iodide: experiment and
  theory},}\ }\href
  {https://doi.org/https://doi.org/10.1016/j.chemphys.2004.05.011} {\bibfield
  {journal} {\bibinfo  {journal} {Chem. Phys.}\ }\textbf {\bibinfo {volume}
  {303}},\ \bibinfo {pages} {179--187} (\bibinfo {year} {2004})}\BibitemShut
  {NoStop}%
\bibitem [{\citenamefont {Yuwono}\ \emph {et~al.}(2024)\citenamefont {Yuwono},
  \citenamefont {Li}, \citenamefont {Zhang}, \citenamefont {Surjuse},
  \citenamefont {Valeev}, \citenamefont {Li},\ and\ \citenamefont
  {DePrince~III}}]{DePrinceIII24_6521}%
  \BibitemOpen
  \bibfield  {author} {\bibinfo {author} {\bibfnamefont {S.~H.}\ \bibnamefont
  {Yuwono}}, \bibinfo {author} {\bibfnamefont {R.~R.}\ \bibnamefont {Li}},
  \bibinfo {author} {\bibfnamefont {T.}~\bibnamefont {Zhang}}, \bibinfo
  {author} {\bibfnamefont {K.~A.}\ \bibnamefont {Surjuse}}, \bibinfo {author}
  {\bibfnamefont {E.~F.}\ \bibnamefont {Valeev}}, \bibinfo {author}
  {\bibfnamefont {X.}~\bibnamefont {Li}},\ and\ \bibinfo {author}
  {\bibfnamefont {A.~E.}\ \bibnamefont {DePrince~III}},\ }\bibfield  {title}
  {\enquote {\bibinfo {title} {Relativistic coupled cluster with completely
  renormalized and perturbative triples corrections},}\ }\href
  {https://doi.org/10.1021/acs.jpca.4c02583} {\bibfield  {journal} {\bibinfo
  {journal} {J. Phys. Chem. A}\ }\textbf {\bibinfo {volume} {128}},\ \bibinfo
  {pages} {6521--6539} (\bibinfo {year} {2024})}\BibitemShut {NoStop}%
\end{thebibliography}%

\end{document}